\documentclass[11pt]{article}
\usepackage{graphics}
\usepackage{epsfig}

\newcommand{\be}{\begin{eqnarray}}
\newcommand{\ee}{\end{eqnarray}}

\def\ll#1{\left#1}
\def\r#1{\right#1}
\def\fr{\frac{1}{2}}

\def\mref#1{(\ref{#1})}

\def\p{\partial}

\def\bd{\begin{displaymath}}
\def\ed{\end{displaymath}}

\def\ba#1{\begin{array}{#1}}
\def\ea{\end{array}}
\def\nn{\nonumber}
\newfont{\Bbb}{msbm10 scaled 1200}

\begin{document}

\pagestyle{empty}

\begin{center}

{\LARGE\bf High energy semiclassical wave functions in rational multi-connected and other (Sinai-like) billiards determined by their periodic orbits\\[0.5cm]}

\vskip 12pt

{\large {\bf Stefan Giller}}

\vskip 3pt

Jan D{\l}ugosz University in Czestochowa\\
Institute of Physics\\
Armii Krajowej 13/15, 42-200 Czestochowa, Poland\\
e-mail: stefan.giller@ajd.czest.pl
\end{center}

\vspace {40pt}

\begin{flushright}
{\it Mojej \.Zonie}
\end{flushright}

\vspace{20pt}

\begin{abstract}
The methods of the high energy semiclassical quantization in the rational polygon billiards used in our earlier papers are generalized to an arbitrary
rational multi-connected polygon billiards i.e. to the billiards which is a rational polygon with other rational polygons inside them "rotated" with
respect to the "mother" ones by rational angles. The respective procedure is described fully and its most important aspects are discussed. This
generalization allows us to apply the method to arbitrary billiards with curved boundaries and with multi-connected areas
where the respective semiclassical quantization is determined by the shortest periodic orbits of the  billiards. As an example of the latter case the
Sinai-like billiards is considered which is the right angle triangle with one of its acute angles equal to $\pi/6$ and with the circular hole in it.
\end{abstract}

\vskip 50pt
\begin{tabular}{l}
{\small PACS number(s): 03.65.-w, 03.65.Sq, 02.30.Jr, 02.30.Lt, 02.30.mZ} \\[1mm]
{\small classical periodic trajectories, chaotic dynamics, quantum chaos, scars, superscars}
\end{tabular}

\newpage

\pagestyle{plain}

\setcounter{page}{1}

\section{Introduction}

\hskip+2em In several of our previous papers \cite{41}-\cite{54} we have developed a way of constructing semiclassical wave functions (SWF) in polygon billiards
both rational (RPB) and
irrational (IPB) ones as well as we have applied the method to the semiclassical quantization of chaotic billiards exemplified by the Bunimovich stadium \cite{52}.
There is a question however about possibilities of generalizing the method to cover also billiards which are multiconnected, i.e. which have also inner
boundaries, see Fig.1. In the present paper we extend the method to such cases.

Billiards due to its extremely simple classical dynamics are particularly convenient dynamical systems for applying the semiclassical approximations to
describe their quantum behaviour especially if such approximations are constructing according to the Maslov-Fedoriuk approach \cite{4} where the
SWFs are built on classical trajectories. Since in the billiards cases classical trajectories are straight lines broken at the boundaries
according to the optical reflection rule the respective SWFs appear to be plane waves with definite wave lengths propagated along these trajectories and
reflected by the billiards boundaries according to the same optical rule, see \cite{42} and Sec.2. Therefore the stationary state wave functions arise
as a result of interferences of many plane waves reflected in the above way and vanishing by these interferences at billiards boundaries.

In general in billiards with some arbitrary geometrical form of their boundary an explicit performing of such
interferences is not an easy task. It is as such however for simpler billiards forms such as the polygon ones (PB). It is well known that in this cases the
optical rule of reflections of waves by the billiards boundaries
allows us to straighten up segments of trajectories broken by reflections by making mirror images of the considered polygon billiards by its
sides. Making as many as necessary of such mirror reflections one transforms any trajectory of the billiards into straight line on the plane, i.e. the
plane can be covered totally by such mirror reflections of PB and each trajectory of the billiards become a straight line on such a plane. This is what is
called unfolding motions in billiards.

However such an unfolding process does not in general lead just to a single plane covered tightly by all mirror reflections of PB. Such
simple situations appear only for PB which are classically integrable (isosceles triangles, some other right triangles, rectangles) while
for other PB the plane is split into infinitely many planes connected with themselves by polygon billiards sides resembling by its structure
Riemann surfaces known from the complex analysis and because of that named also as a polygon billiards Riemann surface (PBRS) \cite{42}.

In general PBRS does not show any regular structure. The exceptions are PB which form a class of them called rational. The rational polygon billiards
are those all angles of which are rational part of $\pi$. In such cases the respective PBRS become periodic, i.e. they are formed by periodically
shifted pattern of a finite number of billiards emages called elementary polygon pattern (EPP), see \cite{42} and Sec.2. The latter arises as a finite number of mirror
reflections of the original RPB and its mirror images in such a way that it contains all different images of the original RPB.

This fact that PBRS corresponding to RPB can be got by periodic shifting of EPP mean that the latter must provide us with a set of periods - these ones
which can be used to get PBRS. In fact the main property of each EPP is that its boundary is built by pairs of parallel sides of the respective RPB
being elements of EPP. The sides
in such pairs belong to two RPB which are mirror images of each other in such sides. The sides making these pairs can be therefore identified making in
this way of EPP a closed two dimensional surface. It is clear therefore that a vector linking any pair of identified in this way points of the sides is
a period. It should be also clear that each such a period coincides with an unfolded periodic orbit of RPB considered. A set of all periods obtained in
this way determines a vector space with integers as coefficients which contains of course a number of linear independent ones.

Each EPP with pairwise identified parallel boundary sides making of it two dimensional closed surface is topologically equivalent to a multitorus of
a genus $g$, see Sec.2. The latter means that there are $2g$ linearly independent periods among those mentioned in the previous paragraph. Because of this
correspondence between EPP and a multitorus of genus $g$ the classical dynamics in RPB is called pseudointegrable if $g>1$ \cite{2}.

If now one is trying to build SWF for pseudointegrable RPB one is met with the problem of finding a function defined on RPRS
on which it should be multiperiodic with more than two periods \cite{42}. In general such functions do not exist since on the level of periods it means satisfying
too many conditions independently.

To be a little bit more precise let us consider the integrable RPB with $g=1$ for which the semiclassical quantization can be done conventionally. In such
a case of RPB it means that there are two linear independent periods of EPP say ${\bf D}_1,{\bf D}_2$ for which one can write the following quantization
conditions
\be
{\bf p}\cdot{\bf D}_1=2\pi m \;\;\;\;\;\;\;\;\;\;\;\;\;\;\;\;\;\;\;\;\;\;\;\;\;\;\;\;\;\;\;\;\;\;\nn\\
{\bf p}\cdot{\bf D}_2=2\pi n, \;\;\;\;\;\;\;\;m,n=0,1,2,...
\label{1}
\ee
which expresses the periodicity of the plane SWF with a momentum {\bf p}.

Since any other period of EPP for the case considered is linear dependent on the two used in \mref{1} (with integer coefficients) then the periodicity
of SWF for this period is satisfied also. It is not the case in general when $g>1$ since then other periods cannot be expressed linearly by the two ones
${\bf D}_1,{\bf D}_2$ with integer coefficients so that the conventional semiclassical quantization of pseudointegrable PB of course fails in general
in such cases. Nevertheless even in these pseudointegrable cases
there is a class of RPB which permits conventional semiclassical quantization. These are so called doubly rational polygon billiards (DRPB) \cite{42} for which any
of its period can always be represented by a linear combination of ${\bf D}_1,{\bf D}_2$ with rational coefficients, i.e. we have
\be
{\bf D}_k=\frac{p_{k1}}{q_{k1}}{\bf D}_1+\frac{p_{k2}}{q_{k2}}{\bf D}_2,\;\;\;\;\;\;k=3,...,2g
\label{2a}
\ee
so that the conditions can be rewritten as
\be
{\bf p}\cdot{\bf D}_1=2\pi Z_1m \;\;\;\;\;\;\;\;\;\;\;\;\;\;\;\;\;\;\;\;\;\;\;\;\;\;\;\;\;\;\;\;\;\;\nn\\
{\bf p}\cdot{\bf D}_2=2\pi Z_2n, \;\;\;\;\;\;\;\;m,n=0,1,2,...
\label{2b}
\ee
where integers $Z_1,Z_2$ are the least common multiples of the denominators $q_{k1}$ and $q_{k2},\;k=3,...,2g$, respectively.

When however the coefficients in \mref{2a} are irrational we can still continue the conventional way of semiclassical quantization of the respective RPB
approximating irrational coefficients in the relations \mref{2a} by corresponding rationals. This however must be done in some sophisticated way to achieve
desired accuracy of the final results. A respective tool for realizing such a goal is provided by the Dirichlet approximation theorem (DAT) \cite{1}. This
theorem is the key one if one wants to extend the conventional semiclassical quantization also on other types of billiards, i.e. on the irrational
polygon billiards (IPB) and on billiards with arbitrary shapes of their boundaries.

However before presenting DAT let us discuss shortly strictly related with the theorem the exceptional role played by periodic orbits
in the semiclassical quantization
presented in the papers  \cite{41}-\cite{54} and in the present one particularly if arbitrary billiards are quantized. This role follows directly from
the well known condition that the semiclassical approximation can be applied essentially
to a quantum system in its high energy regime. In the case of the stationary states of such systems it means that such states can be considered as a
superposition of standing plane waves which wave lengths are respectively short. These waves propagate along classical trajectories and if the
trajectories are periodic the waves must satisfy the conditions of being unambiguous, i.e. the waves must be periodic on periodic trajectories which is
expressed by the conditions like \mref{1}-\mref{2b}. This
means further that a number of wave lengths distributed
along any periodic trajectory must be integer. Therefore the wave lengths of standing waves can be considered as a kind of length measure units by which
lengths of periodic orbits are expressed by integers.

However it is clear
that having many periodic orbits rather independent of each other one cannot expect that their lengths can be measured with a single wave length just
because the orbits can be simply incommensurable. In such cases they can be measured by a given wave length only approximately and the question arises
whether it is possible at all to fix the respective length of the wave in such a way to measure lengths of a set of periodic orbits with an accuracy
being a desired small fraction of the wave length itself. This is just DAT which tells us that such a possibility
does exist and the success in getting of our results of the previous papers and the present one is owed to this theorem.

In the wider context of the measuring theory of physical quantities the Dirichlet approximation theorem provides us with the basic theoretical tool
for such a
measuring. Namely since each physical act of measuring provides us with rational values of such measure independently of whether the measured quantities
are commensurable or not DAT tell us that it is always possible to find a respective unit measure which ensures that inaccuracies which must accompany
each process of measuring can be done less than any given part of the unit used for measuring each quantity of their set considered.

In the mathematical language the Dirichlet approximation theorem says that for any set $A=\{\alpha_1,...,\alpha_p\}$ of real numbers and for any natural
$N$ one can find a natural $Z\leq N$ such that taking the $1/Z$ part of $1$ one can "measure" with such a unit each number of $A$ with an accuracy not
worse than $1/N^\frac{1}{p}$ part of $1/Z$, i.e. we have
\be
\ll|\alpha_k-Z_k\frac{1}{Z}\r |<\frac{1}{N^\frac{1}{p}}\frac{1}{Z},\;\;\;\;\;\;k=1,...,n
\label{1a}
\ee
where $Z_k,\;k=1,...,n$, are all integer.

The following comments to DAT are worth to be done.

\begin{itemize}
\item For a given $N$ a number $Z$ in the condition \mref{1a} is not unique, i.e. for each $N$ there is a set $Z_{min}<...<Z_{max}$
of such numbers $Z$ satisfying the conditions \mref{1a}. Further in the paper a number $Z$ will mean any fixed number of this set.
\item For a given $Z,\;Z_{min}\leq Z\leq Z_{max}$, consider all rationals $Z_k/Z,\;k=1,...,n$, approximating respective $\alpha_k,\;k=1,...,n$. Then
$Z$ is, obviously, the least common multiple for these rationals.
\item The arbitrariness of $N$ in DAT should be meant literally. However the smaller $N$ the worse are the respective approximations in \mref{1a}. In the extreme
case when $N=1$ then $Z=1$ and all $Z_i$ in \mref{1a} are then equal to the integer parts of approximated real numbers, i.e. the respective approximations
are very crude.
\item If all $\alpha_k$ are rational, i.e. $\alpha_k=p_k/q_k,\;k=1,...,n$, with coprime integers $p_k,q_k$ and $C$ denotes the least
common multiple of $q_k,\;k=1,...,n$, then for $N<C$ the theorem has exactly the form as in \mref{1a} while for $N\geq C$ we have $Z_{min}\leq C\leq Z_{max}$
and putting $Z=C$ in \mref{1a} causes the l.h.s. of it vanishing.
\end{itemize}

\begin{figure}
\begin{center}
\psfig{figure=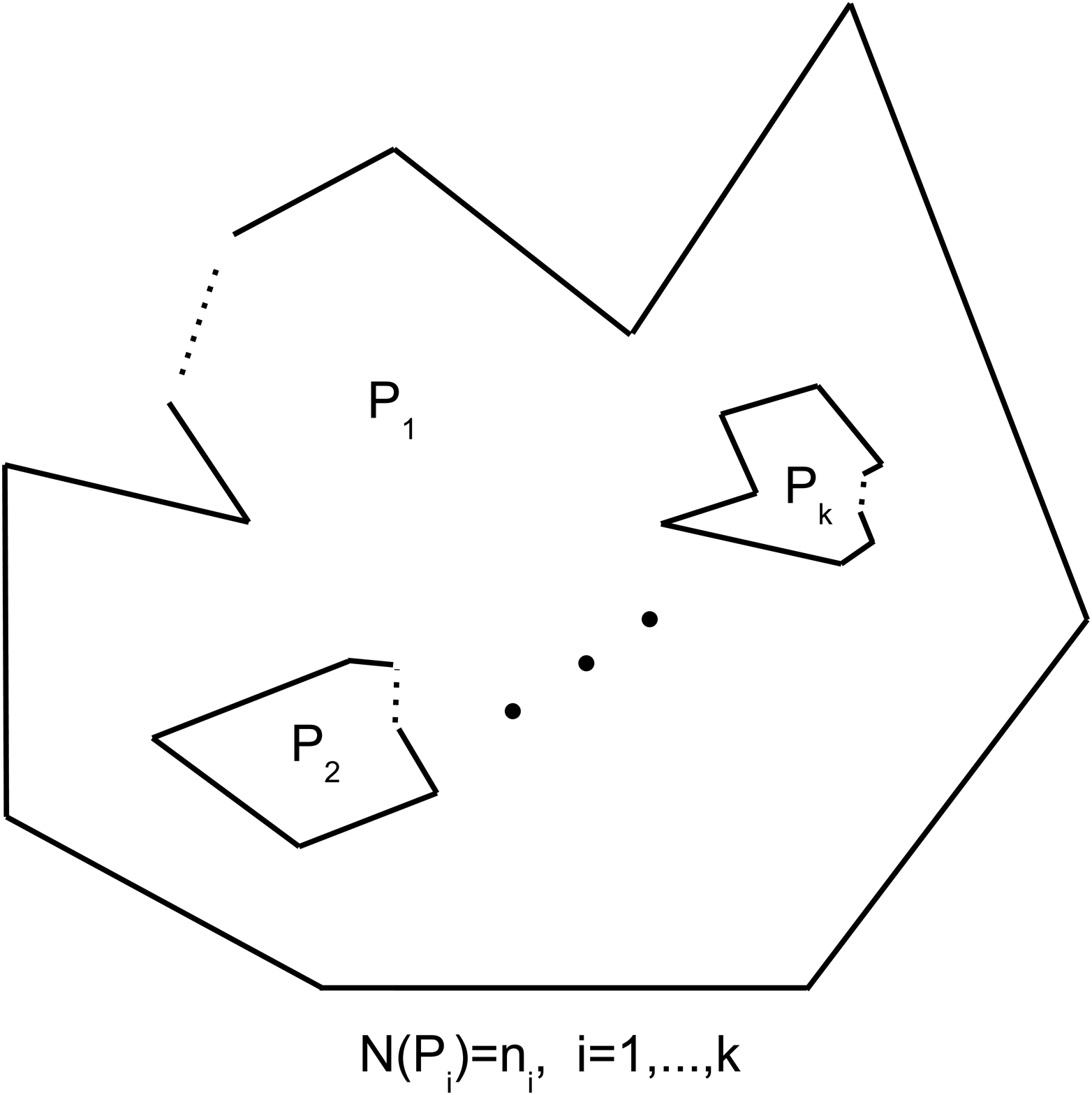,width =13 cm}
\caption{An arbitrary multi-polygon billiards with $(k-1)$-polygon holes. A function $N(P_i)$ provide us with a number $n_i$ of sides in the respective
polygons $P_i,\;i=1,...,k$.}
\end{center}
\end{figure}

The DAT can be applied to rationalize the relations between periods of RPB as well as to rationalize angles in the case of IPB. The first case will be
considered explicitly in the next sections. Considering the second case if $\alpha_k$ in \mref{1a} denote angles of Fig.2 which are now assumed to be
any real numbers then rationalizing them by DAT's we get
\be
\ll|\sum_{j=1}^k\sum_{i=1}^{n_j}\alpha_{ij}^{rat}-\sum_{i=1}^kn_i-2k+4\r|<\frac{\sum_{i=1}^kn_i}{ZN^\frac{1}{\sum_{i=1}^kn_i+k-1}}
\label{1b}
\ee
where $\alpha_{ij}^{rat}=Z_{ij}/Z$ denotes rationalized $\alpha_{ij}$.

The above rationalization of IPB can destroy its polygon form. To maintain it one can rationalize its single angle say $\alpha_{11}$ using the relation
(6) of Fig.2 by
\be
\alpha_{11}^{rat}=\sum_{i=1}^kn_i-2k+4-\sum_{j=2}^k\sum_{i=2}^{n_j}\alpha_{ij}^{rat}-\sum_{j=2}^k\alpha_{1j}^{rat}-\sum_{i=2}^{n_1}\alpha_{i1}^{rat}
\label{1c}
\ee

In such a rationalization of $\alpha_{11}$ the accuracy of it is the following
\be
\ll|\alpha_{11}-\alpha_{11}^{rat}\r|<\frac{\sum_{i=1}^kn_i-1}{ZN^\frac{1}{\sum_{i=1}^kn_i+k-2}}
\label{1d}
\ee

The applications of the Dirichlet theorem to the several cases of polygon billiards and to the Bunimovich one have been demonstrated in our previous
papers \cite{42}-\cite{54} together with the exceptional role played by periodic orbits in the semiclassical quantization of the billiards systems
deprived of inner boundaries (holes). In the present paper we are going to show that the results of these papers and the methods used there can be
generalized directly to arbitrary billiards.

It is important to stress that the approximations mentioned above which have to be done on the respective steps of constructing of the semiclassical
wave functions as well as the energy spectra are controlled by the respective theorems which can be found in the well known monograph of Courant and Hilbert \cite{40}

The paper is organized as follows.

In the next section a necessary and complete resum{\'e} on the semiclassical approach applied in the paper is given standardizing all necessary steps
in extending the approach to any (with holes) rational billiards allowing for building SWFs in such billiards and providing a general formula for
them.

Sections 3. and 4. provide examples of applications of the semiclassical tools developed in Sec.2.

In section 3. two examples of the rational billiards are considered in details and their respective semiclassical wave functions and energy spectra are
constructed and discussed.

In section 4. the case of chaotic billiards is investigated which is the Sinai-like one \cite{3} build of the right triangle with a circle hole in it. Its shortest
periodic orbits are used to approximate it by rational multi-connected polygon billiards (RM-CPB) which next is quantized semiclassically.

The paper is summarized in section 5.

\section{A resum{\'e} of the high energy semiclassical quantization in rational polygon billiards and its extension to the rational multi-polygon billiards}

\begin{figure}
\begin{center}
\psfig{figure=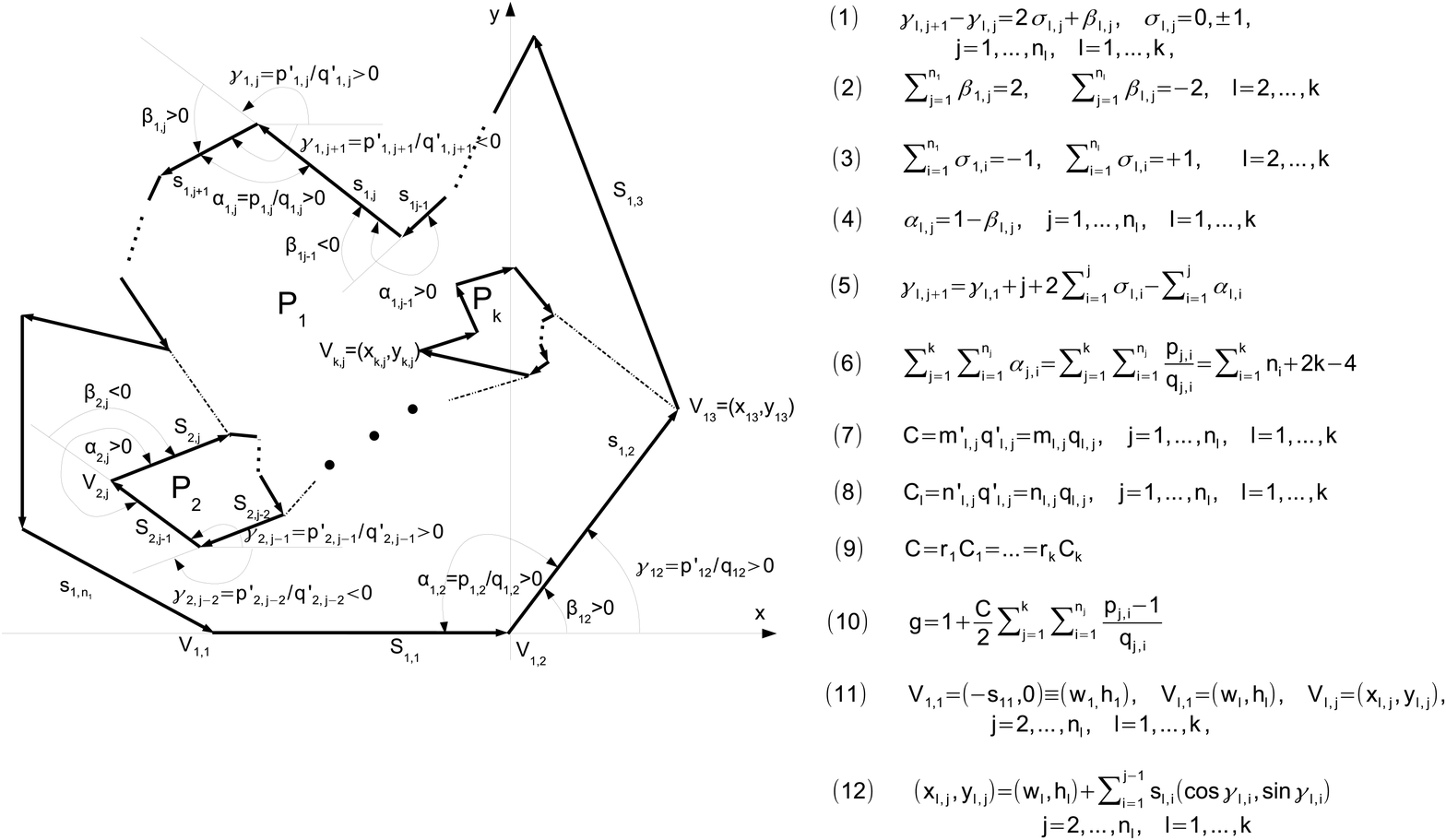,width =10 cm}
\end{center}
\end{figure}
\begin{figure}
\begin{center}
\caption{An arbitrary RM-CPB with $k-1$-polygon holes. All angles are measured by $\pi$-units.
The number $C$ is the least common multiple of {\it all} denominators
of the angles $\gamma_{l,j},\;j=1,...,n_l,\;l=1,...,k$, shown in the figure. The numbers $C_i,\;i=1,...,k$, are the least common multiples of the
denominators of the $\gamma$-angles belonging to the respective polygon $P_i,\;i=1,...,k$. The number $g$ is the genus of a multitorus in the phase space corresponding to the classical motion in the
billiards. The dotted-dash lines linking $2k$ vertexes of the billiards divide its surface into two connected pieces.}
\end{center}
\end{figure}

\subsection{The classical motion in rational multi-connected polygon billiards}

\hskip+2em In the following considerations it is assumed that a multi-connected polygon billiards is rationalized according to methods described in Introduction.

A general type of RM-CPB we are going to consider is shown in Fig.2. Its classical motion is pseudointegrable \cite{2}, i.e. in its phase
space all trajectories of the billiards ball with a given energy lie on a multitorus with a genus $g$ corresponding to the $g$ holes it possesses. This
multitorus has its plane form called elementary polygon pattern (EPP) which contains all $2C$ different mirror reflections of the billiards of Fig.2 in
its sides glued along these sides while pairs of the sides being parts of its boundary are identified. The identified pairs of sides are parallel to
each other and belong to two billiards of EPP which are images of each other when reflected in these sides so that any reflection in any of the boundary
side of EPP leads to a billiard which position repeats some of those forming EPP.

Each EPP formed in this way has periodic structure with each period linking a pair of the same points of two identified parallel sides of the EPP
boundary. Among these periods there are $2g$ ones which are linear independent in the space of integers. Despite that a description of the machinery used
to investigate semiclassically rational billiards by making use of their EPP has been done in details in our previous papers \cite{41}-\cite{53} we have
made its resum{\'e} here extending it to cover also RM-CPB.

\subsection{Forming EPP}

\hskip+1,5 em In our earlier papers \cite{41}-\cite{53} we have described the main property of EPP and build them for the examples of RPB considered in
these papers but we did not give an effective general recipe how to form it. The following provides such a description standardizing it to a large extent.

First as we have mentioned each EPP contains $C$ different positions of the considered RM-CPB together with $C$ of their mirror reflections in the sides of
RM-CPB so that each RM-CPB in EPP is surrounded by its mirror reflections and vice versa. $C$ denotes here the least common multiple of all denominators of the
rational angles of RM-CPB shown in Fig.2. Let us note at this moment that if $C_i,\;i=1,...,k$, denote the least common multiples for the denominators
of angles describing the respective polygons $P_i$ of Fig.2 then $C$ is also the least common multiple for them.

A basic part of EPP can be build beginning with an arbitrary chosen vertex of the billiards. However to standardize its construction let us choose one of the
vertex of the outer polygon billiards of Fig.2 calling it $V_{12}$. Next we make subsequent reflections of the billiards by its
two sides $s_{11}$ and $s_{12}$ forming the vertex $V_{12}$, see Fig.2. If the side $s_{12}$ forms an angle $p'_{12}/q_{12}$ (in the $\pi$-unit) with the
$x$-axis than one can make
$2q_{12}$ such reflections to close the set of different reflections of the billiards obtained in this way. It is to be noticed that a single such a
reflection in one side is a rotation of the other side around the vertex by the angle $2p'_{12}/q_{12}$ while the subsequent reflection in the other side
rotates the billiards by the angle $2p'_{12}/q_{12}$ around the vertex $V_{12}$. There are $q_{12}$ such rotations of the billiards and also $q_{12}$
similar rotations of its odd image which saturates the considered part of EPP. Since the rotations are rational the positions of the sides $s_{11}$ and
$s_{12}$
are distributed uniformly around the vertex $V_{12}$ so that the two neighbor positions of the same side make the angle equal to $2\pi/q_{12}$. The
same angles are made by two closest positions of any other side of the rotated billiards as well as by its odd images, see Fig.3. Of course the part of
EPP got in this way contains $q_{12}$ different positions of RM-CPB and also $q_{12}$ of their mirror reflections. Denote this part by EPP$_{q_{12}}$. It is
a connected surface which is invariant under its rotation by $2\pi/q_{12}$ around the vertex $V_{12}$.

\begin{figure}
\begin{center}
\psfig{figure=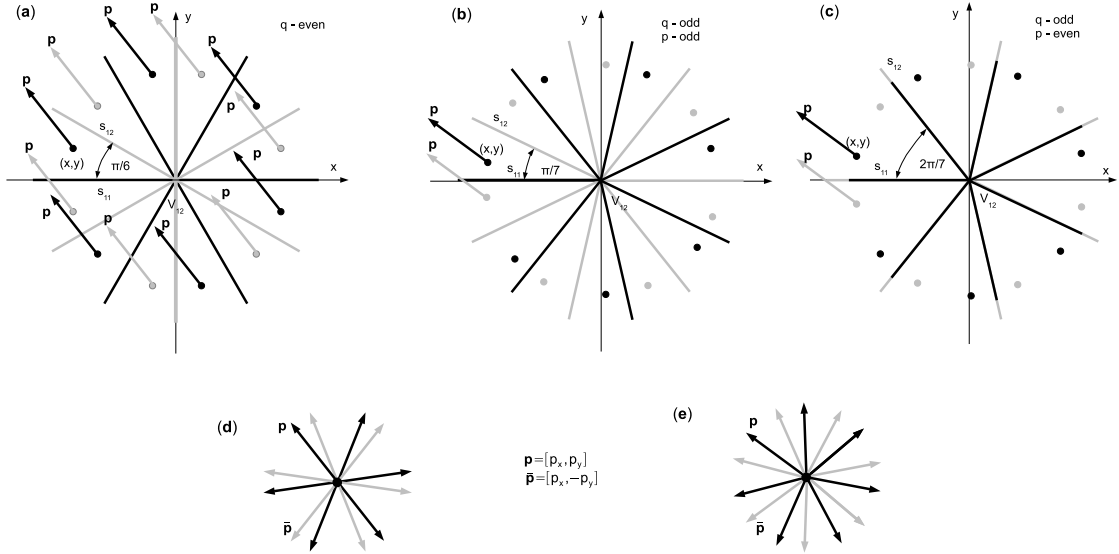,width=15 cm}
\caption{({\bf a})-({\bf c}) - three possibilities of angle distributions of sides $s_{11}$ and $s_{12}$ in EPP$_{q_{12}}$. For any other side $s_{lj}$ the respective
distributions look similarly being rotated by the angle $\pi-\gamma_{lj}$ with respect to the ones on the figure; ({\bf d})-({\bf e}) - distributions
of the skeleton momenta shown in Fig.Fig.({\bf a})-({\bf c}) passing by any point $(x,y)$ of RPB when its EPP$_{q_{12}}$ and the skeleton itself are
folded back to the billiards.}
\end{center}
\end{figure}

Now let us distinguish the following cases in the procedure of forming EPP.
\begin{enumerate}
\item $C=C_1=q_{12}$.

In such a case a forming of EPP is finished on EPP$_{q_{12}}$, i.e. all different positions of each side of RM-CPB of Fig.2 appear in EPP$_{q_{12}}$ in $q_{12}$ different pairs
each of them containing a side of RM-CPB and its reflection parallel to it, i.e. two multi-polygons containing such a twin pair of the same side are
images of each other when reflected
in the side. Of course the sides $s_{11}$ coincide with themselves in their pairs as well as the sides $s_{12}$ while the sides in other pairs are
parallel translations of each other by vectors being the periods of EPP. If $s_{lj}\neq s_{11}, s_{12}$ is some side of RM-CPB of Fig.2 than its twin odd
image can be obtained by its translation by the period ${\bf D}_{lj}^{(1)}$. Since there are $q_{12}-1$ even images $s_{lj}^{(i)},\;i=2,...,q_{12}$ of
$s_{lj}$($\equiv s_{lj}^{(1)}$), arising by subsequent rotations of the
billiards around its vertex $V_{12}$ by the angle $2\pi p'_{12}/q_{12}$ then there also the same number of periods ${\bf D}_{lj}^{(i)},\;i=2,...,q_{12}$
being the rotations of the period
${\bf D}_{lj}^{(1)}$ by the respective angles $2(i-1)\pi/q_{12},\;i=2,...,q_{12}$ and linking these images with their respective twins, i.e. the periods
${\bf D}_{lj}^{(i)},\;i=1,...,q_{12}$ link each side of even images of the billiards (including its original position) with their twin images belonging to
the odd images of the billiards. The total number
of these periods is of course equal to $C(\sum_{l=1}^k\sum_{j=1}^{n_l}n_{lj}-2)$. Among them there are $2g$ independent periods of the multi-torus corresponding
to the case.
\item $C=C_1=n_{12}q_{12},\;n_{12}\neq 1$

In this case there are still other positions of the reflected billiards not represented in EPP$_{q_{12}}$. Nevertheless all $C$ different positions of
any side of the
billiards must be also uniformly distributed inside the angle $2\pi$ making the angle $2\pi/C$ between each pair of the closest
positions of them, i.e. between the closest positions got by forming EPP$_{q_{12}}$ there are still $n_{12}-1$ new positions of the side. Therefore one
can get these new positions and finally EPP itself by rotating EPP$_{q_{12}}$ $(n_{12}-1)$-times around the vertex $V_{12}$ by the angle $2\pi/C$. This
provides us with $n_{12}-1$ new forms EPP$_{q_{12}}^{(u)},\;u=2,...,n_{12}$ (EPP$_{q_{12}}^{(1)}\equiv$EPP$_{q_{12}}$) of EPP$_{q_{12}}$.

If now the odd twin image of the side $s_{lj}(\equiv s_{lj}^{(11)}$, now $s_{lj}^{(uv)}$ denotes the even image of $s_{lj}$ in EPP$_{q_{12}}^{(u)}$ rotated by the
angle $2(v-1)\pi p'_{12}/q_{12},\;u,v=1,...,n_{12}$) is found in some EPP$_{q_{12}}^{(u_{lj})},\;1\leq u_{lj}\leq n_{12}$, then the
odd twin images of {\it all}
the remaining even images of $s_{lj}^{(11)}$ in EPP$_{q_{12}}^{(1)}$ belong to EPP$_{q_{12}}^{(u_{lj})}$ also. Therefore one can follow only the odd
images ${\bar s}_{lj}^{(1u_{lj})}$ of $s_{lj}^{(11)}$ in EPP$_{q_{12}}^{(u_{lj})},\;1\leq u_{lj}\leq n_{12}$, since the remaining ones can be got by
respective rotations, see Fig.4.

\begin{figure}
\begin{center}
\psfig{figure=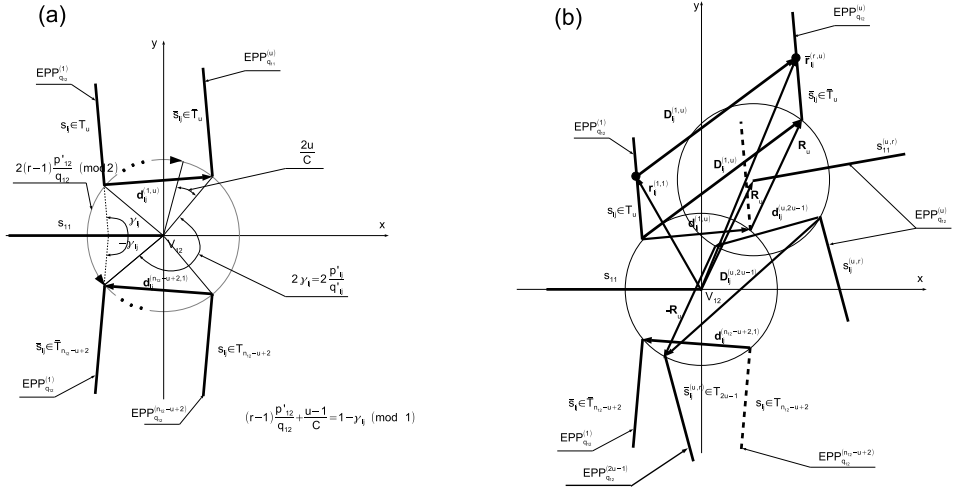,width=13 cm}
\caption{(a) - geometrical relations between positions of sides of RM-CPB in its EPP$_{q_{12}}^{(1)}$ and EPP$_{q_{12}}^{(u)},\;u=2,...,n_{12}$, (b) - the
vectors ${\bf d}_{lj}^{(1,u)}$ and ${\bf d}_{lj}^{(n_{12}-u+2,1)}$ and the respective periods ${\bf D}_{lj}^{(1,u)}$ and
${\bf D}_{lj}^{(n_{12}-u+2,1)},\;u=2,...,n_{12}$ of EPP. All angles are given in $\pi$-units.}
\end{center}
\end{figure}

Therefore for every $u,\;u=1,...,n_{12}$, there is a set $T_u$ of sides $s_{lj}\;(\neq s_{11},s_{12})$ of the multi-polygon billiards {\bf B} of
Fig.2 containing $t_u\geq 0,\;(\sum_{u=1}^{n_{12}}t_u=\sum_{l=1}^k\sum_{j=1}^{n_l}n_{lj}-2)$, of these sides which odd twin images are in the respective
set ${\bar T}_u\subset$ EPP$_{q_{12}}^{(u)}$.

Let ${\bf d}_{lj}^{(1u)},\;u=1,...,n_{12}$ denote translation vectors which
shift $s_{lj}\in T_u$ to its odd twin position  ${\bar s}_{lj}\in {\bar T}_u\subset$EPP$_{q_{12}}^{(u)},\;u=1,...,n_{12}$, see Fig.4. Of course every
${\bf d}_{lj}^{(11)}\equiv{\bf D}_{lj}^{(1)}$ is a period of the final EPP but other ${\bf d}_{lj}^{(1,u)},\;u=2,...,n_{12}$ are not as such. The reason for that is disconnectiveness of the set of
all EPP$_{q_{12}}^{(u)},\;u=1,...,n_{12}$, i.e. they are not glued with themselves.

The situation described above repeats cyclicly. Namely, considering the part EPP$_{q_{12}}^{(2)}$ one detects the sets $T_u^{(2)}$ ($T_u\equiv T_u^{(1)}$) of sides which for every
$u,\;u=1,...,n_{12}$, are respective rotations of the sides of $T_u$ by the angle $2\pi/C$ and the odd images of which are found in the set
${\bar T}_u^{(2)}\subset$EPP$_{q_{12}}^{(u+1)}$. The latter are also the rotation by the angle $2\pi/C$ of the set ${\bar T}_u$.
These odd images are results of shifting each $s_{lj}\in T_u^{(2)}$ by the vectors ${\bf d}_{lj}^{(2,u+1)}$ being the rotations of ${\bf d}_{lj}^{(1u)}$
by the angle $2\pi/C,\;u=1,...,n_{12}$. Again the vectors ${\bf d}_{lj}^{(2,2)}$ are periods while the remaining ones are not.

Obviously for every $v=3,...,n_{12}$, in EPP$_{q_{12}}^{(v)}$ there is set $T_u^{(v)}$ of sides which for every
$u,\;u=1,...,n_{12}$, are respective rotations of the sides of $T_u$ by the angle $2\pi(v-1)/C$ and the odd images of which can be found in
${\bar T}_u^{(v)}\subset$EPP$_{q_{12}}^{(u+v-1)}$ the latter being the rotation by the angle $2\pi(v-1)/C$ of the set ${\bar T}_u$.
The images ${\bar s}_{lj}\in {\bar T}_u^{(v)}$ are results of shifting of each $s_{lj}\in T_u^{(v)}$ by the vectors ${\bf d}_{lj}^{(v,u+v-1)},\;u=1,...,n_{12}$
and again the vectors ${\bf d}_{lj}^{(v,v)}$ are periods while the remaining ones are not. The vectors ${\bf d}_{lj}^{(v,u+v-1)},\;u=1,...,n_{12}$ are
rotations of the vectors ${\bf d}_{lj}^{(1u)}$ by the angle $2\pi(v-1)/C$.

Before proceeding further let us first express the vectors ${\bf d}_{lj}^{(u,v)}$ by coordinates of respective ends of sides connected by these vectors.

According to Fig.2 if $(x,y)$ are coordinates of some point of the billiards then $(x,-y)$ are
coordinates of its image after the first reflection of the billiards in the side $s_{11}$. Coordinates of all other even images of both the point $(x,y)$
and its odd image $(x,-y)$ can be obtained by their respective rotations. Namely, as we have discussed it earlier both the points are
first distributed inside EPP$_{q_{12}}^{(1)}$ by their rotations $r$-times $r=1,...,q_{12}-1$ by the angle $2\pi p'_{12}/q_{12}$ around the point
$V_{12}$ and next each of these points is rotated by the angle $2\pi (u-1)/C$ to find itself in the part EPP$_{q_{12}}^{(u)},\;u=2,...,n_{12}$. Therefore
the respective coordinates of the images of points $(x,y)$ and $(x,-y)$ in EPP$_{q_{12}}^{(u)}$ are
\be
x^{(ru)}=x\cos\alpha_{ru}-y\sin\alpha_{ru}\nn\\
y^{(ru)}=x\sin\alpha_{ru}+y\cos\alpha_{ru}
\label{Ac}
\ee
and
\be
{\bar x}^{(ru)}=x\cos\alpha_{ru}+y\sin\alpha_{ru}\nn\\
{\bar y}^{(ru)}=x\sin\alpha_{ru}-y\cos\alpha_{ru}\nn\\
u=2,...,n_{12}
\label{Ad}
\ee
where $\alpha_{ru}=2\pi (n_{12}p'_{12}(r-1)+u-1)/C,\;r=2,...,q_{12},\;u=2,...,n_{12}$ so that $x^{(11)}={\bar x}^{(11)}=x,\;y^{(11)}=-{\bar y}^{(11)}=y$.

Therefore for the twin image ${\bar s}_{lj}\in {\bar T}_u$ of the side $s_{lj}\in T_u$ in EPP$_{q_{12}}^{(u)}$ we have
\be
{\bar x}_{lj}=x_{lj}\cos(2\pi\gamma_{lj})-y_{lj}\sin(2\pi\gamma_{lj})\nn\\
{\bar y}_{lj}=-x_{lj}\sin(2\pi\gamma_{lj})-y_{lj}\cos(2\pi\gamma_{lj})
\label{Ae}
\ee
while the integers $r,u$ according to Fig.4 are given by the equation
\be
(r-1)\frac{p'_{12}}{q_{12}}+\frac{u-1}{C}=1-\gamma_{lj}\;(mod\; 1)
\label{Af}
\ee

Therefore for the vectors ${\bf d}_{lj}^{(1u)},\;u=1,...,n_{12}$, we get
\be
{\bf d}_{lj}^{(1u)}=-[2x_{lj}\sin^2(\pi\gamma_{lj})+y_{lj}\sin(2\pi\gamma_{lj}),x_{lj}\sin(2\pi\gamma_{lj})+2y_{lj}\cos^2(\pi\gamma_{lj})]
\label{Ag}
\ee

The remaining vectors ${\bf d}_{lj}^{(u,v)}$ can be obtained from the above ones by rotations by the respective angles $\alpha_{ru}$. It is easy
to note that the total number of these vectors is then equal to $C(\sum_{l=1}^k\sum_{j=1}^{n_l}n_{lj}-2)$

Now let us make the set of EPP$_{q_{12}}^{(u)},\;u=1,...,n_{12}$, connected. We can proceed as follows assuming for simplicity that $T_1\neq \emptyset$
and let $s_{l_0j_0}\in T_1$. Then (see Fig.5)
\begin{itemize}
\item translate EPP$_{q_{12}}^{(2)}$ by the vector ${\bf R}_{2}=-{\bf d}_{l_0j_0}^{(12)}$ and glue $s_{l_0j_0}$ in this way with its odd image
${\bar s}_{l_0j_0}$ in ${\bar T}_1\subset$EPP$_{q_{12}}^{(2)}$;
\item translate EPP$_{q_{12}}^{(3)}$ by the vector ${\bf R}_{3}=-{\bf d}_{l_0j_0}^{(12)}-{\bf d}_{l_0j_0}^{(23)}$ and glue the respective
$s_{l_0j_0}\in T_1^{(2)}\subset$EPP$_{q_{12}}^{(2)}$ with its odd image ${\bar s}_{l_0j_0} \in{\bar T}_1^{(2)}\subset$EPP$_{q_{12}}^{(3)}$;
\item repeat the procedure up to EPP$_{q_{12}}^{(n_{12})}$ which after the translation by the vector
${\bf R}_{n_{12}}=-{\bf d}_{l_0j_0}^{(12)}-...-{\bf d}_{l_0j_0}^{(n_{12}-1,n_{12})}$ is to be glued with EPP$_{q_{12}}^{(n_{12}-1)}$ at its side
$s_{l_0j_0}\in T_1^{(n_{12}-1)}$ with the side ${\bar s}_{l_0j_0} \in{\bar T}_1^{(n_{12}-1)}\subset$EPP$_{q_{12}}^{n_{12}}$;
\item construct the periods ${\bf D}_{lj}^{(v,u+v-1)},\;u=1,...,n_{12},\;v=1,...,n_{12}$, of EPP according to
\be
{\bf D}_{lj}^{(v,u+v-1)}={\bf d}_{lj}^{(v,u+v-1)}+{\bf R}_v\nn\\
{\bf R}_1={\bf 0}
\label{Ah}
\ee
\end{itemize}

The total number of periods in the case considered is equal to $C(\sum_{l=1}^k\sum_{j=1}^{n_l}n_{lj}-2)-n_{12}+1$ and is again larger than $2g$ for the case.

\item $C=r_1C_1=r_1n_{12}q_{12}=m_{12}q_{12},\;C\neq C_i,\;i=1,...,k$

Constructionally this case does not differ essentially from the last one and can be obtained from it by substituting in the respective considerations $n_{12}$
by $m_{12}$ giving the number of copies of EPP$_{q_{12}}$ obtained by rotating the latter by the angles $2n\pi/C,\;n=1,...,m_{12}-1$. In particular
the total number of periods ${\bf D}_{lj}^{(v,u)}$ linking the even sides of EPP with their respective twin odd ones is equal to
$C(\sum_{l=1}^k\sum_{j=1}^{n_l}n_{lj}-2)-m_{12}+1$.
\end{enumerate}

\begin{figure}
\begin{center}
\psfig{figure=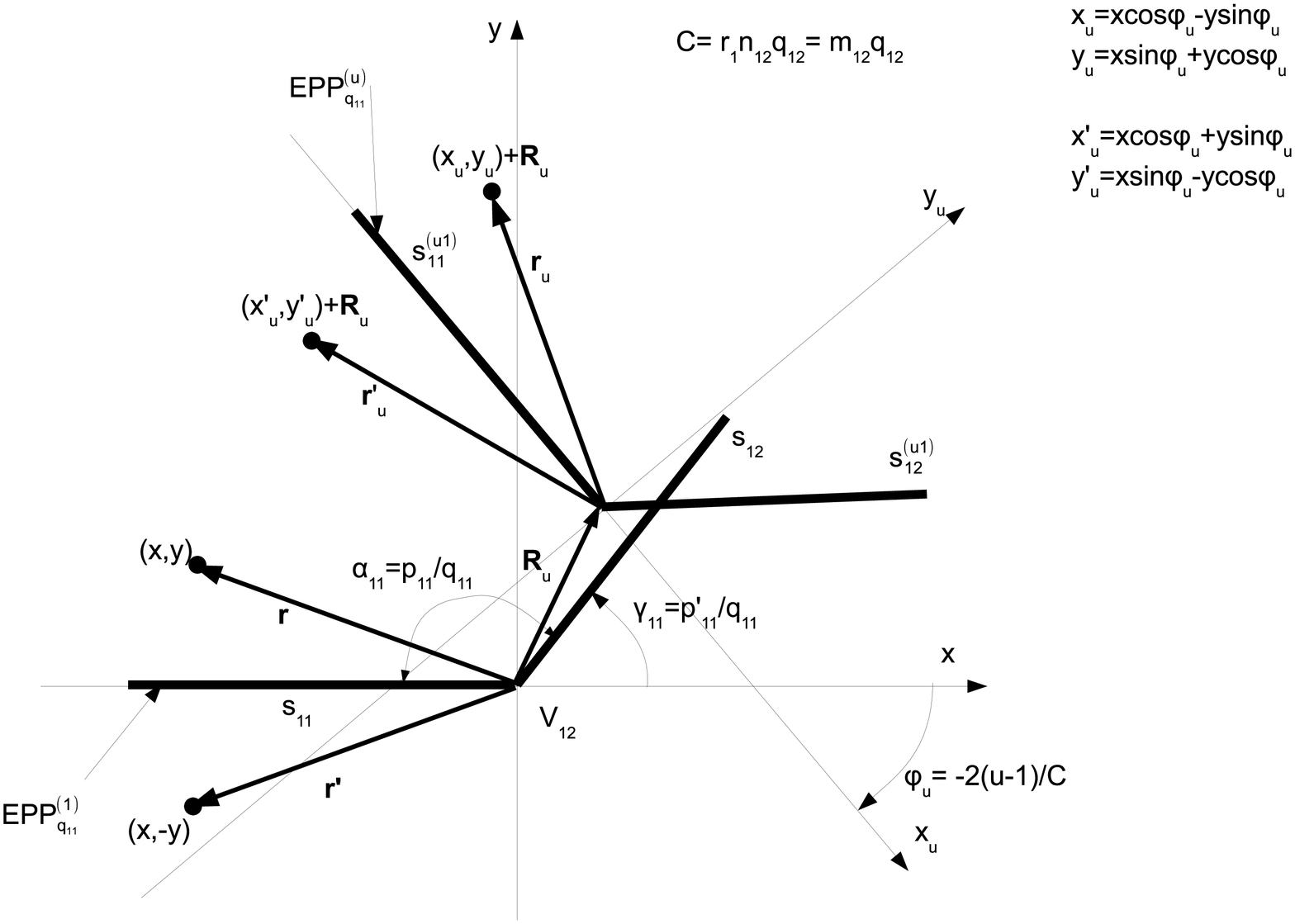,width=9 cm}
\caption{EPP on which SWF is constructed}
\end{center}
\end{figure}

Let us summarize the above way of constructing EPP as follows
\begin{enumerate}
\item choose any vertex $V_{ji}$ of RM-CPB and form around it the corresponding EPP$_{q_{ji}}^{(1)}$;
\item rotate EPP$_{q_{ji}}^{(1)}$ by the angle $2k\pi/C$ to get the respective EPP$_{q_{ji}}^{(k+1)},\;k=1,...,n_i-1$;
\item glue all EPP$_{q_{ji}}^{(k)},\;k=1,...,m_{ji}$ together along $m_{ji}-1$ respective parallel sides making their boundaries - the two
EPP$_{q_{ji}}^{(k)},\;k=1,...,m_{ji}$ glued by such sides must be the mirror reflection of each other; and
\item identify the remaining parallel sides of EPP$_{q_{ji}}^{(k)},\;k=1,...,m_{ji}$ by introducing periods linking them - among the latter there are
$2g$ linearly independent periods corresponding to the multi-torus defined by the constructed EPP.
\end{enumerate}

\subsection{Rationalizing relations between periods}

\hskip+1,5 em Since the periods of EPP link its twin parallel sides they can be identified by respective vertices ending the sides, i.e. as differences of
respective coordinates of these vertices. The coordinates themselves can be given in the coordinate system defined by the chosen two independent
periods, say ${\bf D}_k,\;k=1,2$, in which the system the directions of the periods coincide with the ones of the respective coordinate axes. Let
${\bf D}_{(ij)}={\bf r}_{(j)}-{\bf r}_{(i)}$ be a period defined by two respective vertices with the coordinates ${\bf r}_{(i)}=(x_{(i)},y_{(i)})$ and
${\bf r}_{(j)}=(x_{(j)},y_{(j)})$ in the system mentioned and let the lengths of the chosen periods be taken as the units in their respective directions.
Then we have
\be
{\bf D}_{(ij)}=(x_{(j)}-x_{(i)}){\bf D}_1+(y_{(j)}-y_{(i)}){\bf D}_2
\label{Aa}
\ee

If now the coordinates $(x_{(i)},y_{(i)})$ of the EPP vertices can be represented as the following linear combinations
\be
x_{(i)}=\sum_{k=1}^{n_x}r_{(i),k}\alpha_k\nn\\
y_{(i)}=\sum_{k=1}^{n_y}s_{(i),k}\beta_k
\label{Ab}
\ee
of the sets of real numbers $\alpha_k,\;k=1,...,n_x$, and $\beta_k,\;k=1,...,n_y$ linear independent in the rational spaces of the coefficients
$r_{(i),k},\;s_{(i),k}$ then we can apply DAT to these sets to generate respective integers $N_x,N_y,Z_x,Z_y$ and $C_x,\;C_y$ the latter being the least common
multiples for the rationals  $r_{(i),k},\;s_{(i),k}$ respectively. The integers $n_x,n_y$ defining the respective exponents of $N_x,N_y$ in DAT cannot obviously be then
greater than $C(\sum_{l=1}^k\sum_{j=1}^{n_l}n_{lj}-2)-m_{12}-1$

\subsection{Counting the genus $g$ of the multitorus}

\hskip+1,5 em Realizing the point 4. above we get two dimensional closed (without boundaries) surface which must be equivalent to some multitorus of
genus $g$. Its value is given on Fig.2. While such a result has been obtained by several authors \cite{2}-\cite{6} it is worthwhile to get
it directly for the considered multi-connected billiards. The way of getting it relies on the Euler relation between a geodesic net drawn on the
multitorus and of its genus $g$. Namely, one can easily check that if this net is composed of $S$ simply connected faces boundaries of which are composed of a
finite number $E$ of smooth curves which meet themselves on the multitorus in $V$ different points then we have
\be
E-V-S=2g-2
\label{A0}
\ee

Considering the way by which the respective multitorus is obtained from EPP corresponding to the billiards of Fig.2 and dividing this billiard into
two ones by the dotted-dash lines linking the $k-1$ holes we see that we get in this way the geodesic net for which the relation \mref{A0} is valid.
The respective quantities in \mref{A0} according to Fig.2 are then
\be
E=C\sum_{j=1}^kn_j+2Ck=C\ll(\sum_{j=1}^k\sum_{i=1}^{n_j}\frac{p_{ji}}{q_{ji}}-2k+4\r)+2Ck=C\sum_{j=1}^k\sum_{i=1}^{n_j}\frac{p_{ji}}{q_{ji}}+4C\nn\\
V=\sum_{j=1}^k\sum_{i=1}^{n_j}m_{ji}=C\sum_{j=1}^k\sum_{i=1}^{n_j}\frac{1}{q_{ji}}\nn\\
S=4C
\label{A0a}
\ee

Substituting the above quantities to \mref{A0} we get the standard result ($10$) of Fig.2.

\subsection{Constructing BSWF and SWF on EPP}

\hskip+1,5 em Having formed an EPP corresponding to a given rational billiards we can choose a skeleton on the corresponding RM-CPBRS taking a given
classical momentum {\bf p} and start to
quantize the classical motion on it performed with this momentum \cite{42}. The skeleton is then defined on RM-CPBRS by all trajectories parallel to {\bf p}.

\subsubsection{Constructing BSWF}

\hskip+1,5 em If in the Cartesian coordinate system we choose the $y$-coordinate axis directing parallel to the skeleton then the basic SWF (BSWF)
which can be defined on it has the form
\be
\Psi^\sigma(x,y;p)=e^{\sigma ipy}\chi^\sigma(x,y;p),\;\;\;\;\sigma=\pm
\label{A1}
\ee
where $p=|{\bf p}|$ and $\chi^\sigma(x,y;p)$ satisfies the following Schr{\"o}dinger equation
\be
\sigma 2ip\frac{\p\chi^\sigma(x,y;p)}{\p y}+\ll(\frac{\p^2}{\p {x}^2}+\frac{\p^2}{\p {y}^2}\r)\chi^\sigma(x,y;p)+
(2E-p^2)\chi^\sigma(x,y;p)=0
\label{A2}
\ee
where $E$ is the energy parameter.

In the semiclassical limit $p\to\infty$ the factor $\chi^\sigma(x,y;p)$ is given by the following semiclassical series
\be
\chi^\sigma(x,y;p)=\sum_{k\geq 0}\frac{\chi_k^\sigma(x,y)}{p^k}
\label{A3}
\ee
while the quantized energy $E$ is looked for in the form
\be
E=\frac{1}{2}p^2+\sum_{k\geq 0}\frac{E_k}{p^k}
\label{A4}
\ee

The coefficients $\chi_k^\sigma(x,y)$ of the expansion \mref{A3} satisfy the following recurrent relations
\be
\chi_0^\sigma(x,y)\equiv\chi_0^\sigma(x)\nn\\
\chi_{k+1}^\sigma(x,y)=\chi_{k+1}^\sigma(x)+
\frac{\sigma i}{2}\int_0^{y}\ll(\ll(\frac{\p^2}{\p {x}^2}+\frac{\p^2}{\p z^2}\r)\chi_{k}^\sigma(x,z)+
2\sum_{l=0}^kE_{k-l}\chi_l^\sigma(x,z)\r)dz\nn\\
k=0,1,2,...,
\label{A5}
\ee

As it was shown in our earlier papers \cite{41} two general forms of the solutions to \mref{A5} satisfying the demands of periodicity on the chosen two
periods ${\bf D}_1,{\bf D}_2$ linear independent on the plane are
\begin{enumerate}
\item if none of the periods of $\Psi^\sigma(x,y;p)$ is parallel to the $y$-axes then
\be
\chi_0^\sigma(x)\equiv const\neq 0\;\;\;\;\;\;\;\;\;\;\;\;\nn\\
\chi_k^\sigma(x,y)\equiv 0,\;\;\;\;\;\;\;\;\;\;\;\;k\geq 1
\label{A6}
\ee
while the momentum {\bf p} satisfies the following quantization conditions
\be
{\bf p}\cdot{\bf D}_1=pD_{1y}=2m\pi Z_1\nn\\
{\bf p}\cdot{\bf D}_2=pD_{2y}=2n\pi Z_2\nn\\
m,n\geq \pm 1,\pm 2, ...
\label{A5a}
\ee
where $Z_1=Z_xC_x$ and $Z_2=Z_yC_y$.

The conditions \mref{A5a} have the following solution
\be
{\bf p}_{mn}=2\pi\frac{(Z_1m{\bf D}_2-Z_2n{\bf D}_1)\times({\bf D}_1\times{\bf D}_2)}{|{\bf D}_1\times{\bf D}_2|^2}\nn\\
m,n\geq \pm 1,\pm 2, ...
\label{A5b}
\ee
with the following energy levels
\be
E_{mn}=\frac{1}{2}{\bf p}_{mn}^2=2\pi^2\frac{m^2Z_1^2D_2^2-2mnZ_1Z_2{\bf D}_1\cdot{\bf D}_2+n^2Z_2^2D_1^2}{|{\bf D}_1\times{\bf D}_2|^2}\nn\\
m,n\geq \pm 1,\pm 2, ...
\label{A6a}
\ee

The last formula for energy levels takes the form
\be
E_{mn}=2\pi^2\ll(\frac{m^2Z_1^2}{D_{1}^2}+\frac{n^2Z_2^2}{D_{2}^2}\r)\nn\\
m,n\geq \pm 1,\pm 2, ...
\label{A6c}
\ee
if the periods ${\bf D}_1$ and ${\bf D}_2$ with respective lengths $D_1$ and $D_2$ are orthogonal to each other.
\item Changing the pair ${\bf D}_1,{\bf D}_2$ of periods to another pair of them say ${\bf D}_3,{\bf D}_4$ related with the previous one
by
\be
{\bf D}_i=a_{i1}{\bf D}_1+a_{i2}{\bf D}_2,\;\;\;\;\;i=3,4
\label{A6e}
\ee
provides us with another series \mref{A6a} of energy levels. Nevertheless if both the energy spectra have some common part of energies then
there is an infinite number of them being close to each other
with well defined accuracy and which can even coincide if the coefficients in \mref{A6e} are rational;
\item If there is a period ${\bf D}\approx\frac{Z_{1D}}{Z_1}{\bf D}_1+\frac{Z_{2D}}{Z_2}{\bf D}_2$ parallel to the $y$-axes then there is at least one
POC being a component of the skeleton considered which periodic trajectories has {\bf D} as their period and inside such a POC we have
\be
\chi_0^\sigma(x)=A_\sigma e^{i\sqrt{2E_0}x}+B_\sigma e^{-i\sqrt{2E_0}x}\;\;\;\;\;\;\;\;\;\;\;\;\nn\\
\chi_k^\sigma(x,y)\equiv 0,\;\;\;\;\;\;\;\;\;\;\;\;k\geq 1
\label{A6d}
\ee
with the following conditions on $E_0$, $p$ and the periods ${\bf D}_1$ and ${\bf D}_2$
\be
\begin{array}{lr}
(a)&kZ_1D_{2y}=lZ_2 D_{1y}\\
(b)&rZ_1D_{2x}=sZ_2 D_{1x}\\
(c)&p_nD_{1y}=2\pi nkZ_1\\
(d)&\sqrt{2E_{0,m}}D_{1x}=2\pi mrZ_1\\
(e)&E_{mn}=\frac{1}{2}p_n^2+E_{0,m}=2\pi^2\ll(\frac{m^2r^2Z_1^2}{D_{1x}^2}+\frac{n^2k^2Z_1^2}{D_{1y}^2}\r)\\
(f)&E_0<<\frac{1}{2}p^2\\
&m,n=0,\pm 1,\pm 2,...
\end{array}
\label{A7}
\ee
where the condition $(a)$ and $(b)$ are to be satisfied for some pairs of coprime integers $k,l,\;|k|+|l|>0$, and $r,s,\;|r|+|s|>0,$ respectively.

Defining quantum momenta ${\bf p}_{\pm,m}^{cor}$ and ${\bf p}_{\pm,mn}^{(q)}$ by \cite{42}
\be
{\bf p}_{\pm,m}^{cor}=\ll[\pm\sqrt{2E_{0,m}},0\r]\nn\\
{\bf p}_{mn}^{(q)}={\bf p}_n+{\bf p}_{\pm,m}^{cor}=\ll[\pm{2E_{0,m}},p_n\r]
\label{A7a}
\ee
we can write the solutions to \mref{A7} in the following forms similar to \mref{A5b}
\be
{\bf p}_{n}=2\pi n\frac{(kZ_1{\bf D}_2-lZ_2{\bf D}_1)\times({\bf D}_1\times{\bf D}_2)}{|{\bf D}_1\times{\bf D}_2|^2}\nn\\
{\bf p}_{\pm,m}^{cor}=\pm 2\pi m\frac{(rZ_1{\bf D}_2-sZ_2{\bf D}_1)\times({\bf D}_1\times{\bf D}_2)}{|{\bf D}_1\times{\bf D}_2|^2}\nn\\
{\bf p}_{\pm,mn}=2\pi\frac{((\pm mr+nk)Z_1{\bf D}_2-(\pm ms+nl)Z_2{\bf D}_1)\times({\bf D}_1\times{\bf D}_2)}{|{\bf D}_1\times{\bf D}_2|^2}\nn\\
m,n\geq \pm 1,\pm 2, ...
\label{A5b}
\ee
and
\be
E_{mn}=\frac{1}{2}\ll({\bf p}_{\pm,mn}^{(q)}\r)^2=2\pi^2\times\nn\\
\frac{(\pm mr+nk)^2Z_1^2D_2^2-2(\pm mr+nk)(\pm ms+nl)Z_1Z_2{\bf D}_1\cdot{\bf D}_2+(\pm ms+nl)^2Z_2^2D_1^2}{|{\bf D}_1\times{\bf D}_2|^2}\nn\\
\label{A7c}
\ee

The corresponding BSWF takes then the form
\be
\Psi_{mn}^\sigma(x,y;p)=A_{+,mn}^\sigma e^{\sigma i{\bf p}_{+,mn}^{(q)}{\bf r}}+A_{-,mn}^\sigma e^{\sigma i{\bf p}_{-,mn}^{(q)}{\bf r}},\;\;\;\;\sigma=\pm
\label{A7b}
\ee

It is seen also from the conditions \mref{A7} that the ratios $D_{1x}/ D_{2x}$ and $D_{1y}/ D_{2y}$ or their inversions must be rationalized if the
conditions are to be consistent. If it is done than from the conditions $(a)$ and $(b)$ one can read out the values of the respective pairs $k,l$ and
$r,s$.

The latter remark however is no longer valid if one of the periods say ${\bf D}_1$ is orthogonal to the $y$-axes while the second one ${\bf D}_2$ - to the
$x$-axes, i.e. when ${\bf D}_2$ defines the periodic skeleton and both the periods are orthogonal to each other, since then the conditions are satisfied
for the pair $k=0,l=1$ in the case $(a)$ and for the pair $r=1,s=0$ in the case $(b)$ of the conditions
\mref{A7}. In such a case the energy levels are given by
\be
E_{mn}=2\pi^2\ll(\frac{m^2Z_1^2}{D_{1}^2}+\frac{n^2Z_2^2}{D_{2}^2}\r)
\label{A8}
\ee
where $D_1$ and $D_2$ are the lengths of the respective periods so that the last formula coincides in its form with the formula \mref{A6c} despite of
that the latter one describes energy levels in aperiodic skeletons.

The condition $(f)$ in \mref{A7} is of cause the natural consequence of that the series \mref{A3}-\mref{A4} are the asymptotic ones. However since both the series
are finite then this restriction may appear to be not very essential in many cases.
\item If the skeleton considered is composed of POCs only, i.e. it contains none aperiodic trajectory then the global SWF has the form \mref{A7b} while
it has the form $e^{i{\bf p}\cdot{\bf r}}$ in the opposite case, i.e. if there are aperiodic trajectories in the skeleton.
\item The difference between the forms of SWFs in the two above cases of the skeletons can be interpreted as different results of diffractions
of plane waves propagating along the skeletons by RM-CPB vertexes of which cover the Riemann surface of RM-CPB (RM-CPBRS) periodically.
\begin{itemize}
\item An aperiodic skeleton is composed of trajectories none of which meets
any vertex of RM-CPBRS while each singular trajectory of such a skeleton meets on its way a single vertex only. Let us project a set of all such vertexes
on a segment of straight line by which the aperiodic skeleton considered is crossed orthogonally. Let us enumerate also the skeleton
vertexes according to their increasing distances from the segment by negative integers in the back direction of the skeleton and by positive ones in its
forward direction. Distances of the vertices to the lines are of course definite but irregular changing almost "chaotically". Then the set of points of
the mentioned projection on the segment is dens, countable but the
points themselves are also "chaotically" distributed along the segment if the enumeration of them is followed. Each such a point of the segment is
crossed by a singular trajectory of the skeleton while its remaining points are crossed by the skeleton trajectories. A plane wave propagating by the
aperiodic skeleton is diffracted by met vertexes and since such diffractions happen on the different and "random" distances from the chosen line then
the diffractive  waves interfere with "chaotic" phases which must act destructively in the non forward directions, i.e. such a diffractions is
similar to the scattering by a diffractive grating with chaotically distributed slots and with phases of the diffracted waves depending also
"chaotically" on a slot. It is clear that the corresponding diffractive figure
is again a plane wave still running in the same direction, i.e. along the skeleton.

\item In the case of a periodic skeleton a plane wave running through any of its POCs parallel to it does not meet any vertex. The latter occupy only the singular
diagonals of the POC in a regular way controlled by a period of POC. Therefore the plane wave can be scattered only on the singular diagonal vertexes
with a well defined
diffractive figure, i.e. producing component plane waves with momenta ${\bf p}_\pm^{cor}$ perpendicular to the POC itself. Since the lengths of the
propagating waves are short in the semiclassical approximation considered then these diffractive effects are small and their quantities are
controlled by the condition $(f)$ in \mref{A7}. Such an interpretation of the POC wave propagation as described by \mref{A7}-\mref{A7b} differs essentially
from this given by Bogomolny and Schmit \cite{46}.

\item If the skeleton is composed of POCs only a diffractive picture typical for each POC is maintained by the global skeleton itself with necessary tunings
of the plane waves on singular diagonals.

\item If however the skeleton contains any aperiodic component the chaotic behaviour of the latter in the directions perpendicular to the skeleton destroys
also the plane wave components propagating in these directions in POCs, i.e. the global SWF for the skeleton looks as the latter is totally an aperiodic one.
\end{itemize}
\end{enumerate}

\subsubsection{Constructing SWF on EPP for MPRB}

\hskip+1,5 em The description of EPP in Sec.2.2 allows us to build SWF satisfying the Dirichlet boundary conditions by two steps. First we can build it in
EPP$_{q_{ji}}^{(u)}$ shifted by ${\bf R}_u,\;({\bf R}_1={\bf 0})$, summing next over $u,\;u=1,...,m_{ji}$. Taking into account \mref{Ac}-\mref{Ad} we have
\be
\Psi_{mn}^{(12)}(x,y)=\nn\\
\sum_{u=1}^{m_{12}}(-1)^{u-1}e^{i{\bf p}_{mn}{\bf R}_u}\sum_{r=1}^{q_{12}}\ll(e^{i(p_{mn,x}x^{(ru)}+p_{mn,y}y^{(ru)})}-
                                   e^{i(p_{mn,x}{\bar x}^{(ru)}+p_{mn,y}{\bar y}^{(ru)})}\r)
\label{A9}
\ee

Let us note however that the form of SWF \mref{A9} can be modified by fixing the point $(x,y)$ and summing over all directions of momenta with which
trajectories passes by the point. According to \mref{Ac}-\mref{Ad} we have instead of \mref{A9}
\be
\Psi_{mn}^{(12)}(x,y)=\nn\\
\sum_{u=1}^{m_{12}}(-1)^{u-1}e^{i{\bf p}_{mn}{\bf R}_u}\sum_{r=1}^{q_{12}}\ll(e^{i(p_{mn,x}^{(ru)}x+p_{mn,y}^{(ru)}y)}-
                                   e^{i({\bar p_{mn,x}^{(ru)}}x+{\bar p_{mn,y}^{(ru)}}y)}\r)
\label{A9a}
\ee
where
\be
p_{mn,x}^{(ru)}=p_{mn,x}\cos\alpha_{ru}+p_{mn,y}\sin\alpha_{ru}\nn\\
p_{mn,y}^{(ru)}=-p_{mn,x}\sin\alpha_{ru}+p_{mn,y}\cos\alpha_{ru}
\label{A9b}
\ee
and
\be
{\bar p_{mn,x}^{(ru)}}=p_{mn,x}\cos\alpha_{ru}-p_{mn,y}\sin\alpha_{ru}\nn\\
{\bar p_{mn,y}^{(ru)}}=-p_{mn,x}\sin\alpha_{ru}-p_{mn,y}\cos\alpha_{ru}\nn\\
r=1,...,q_{12},\;\;\;u=1,...,m_{12}
\label{A9c}
\ee

\subsubsection{Accuracies of vanishing of SWF on sides of MPRB}

\hskip+1,5 em  $\Psi_{mn}^{(12)}(x,y)$ by its construction vanishes on the sides $s_{11}$ and $s_{12}$ while on the remaining sides of the billiards
considered it vanishes only approximately as it is shown by the following calculation.

Suppose a point ${\bf r}_{lj}=(x_{lj},y_{lj})$ to lie on the side $s_{lj}\neq s_{11},s_{12}$ of the billiards. Therefore all its even images in EPP lie
also on the even images of $s_{lj}$ in EPP while its odd images lie on the respective odd images ${\bar s}_{lj}$ of $s_{lj}$ in EPP, see Fig.4(b). If
further the twin parallel side of $s_{lj}$ belongs to some of $T_u,\;u=1,2,...,m_{12}$ then all its even images in EPP$_{q_{12}}$ also have their twin
parallel sides in the same EPP$_{q_{12}}^{(u)},\;u=1,2,...,m_{12}$. Therefore the twin parallel images of even images of $s_{lj}$ in
EPP$_{q_{12}}^{(v)}$ all belong to EPP$_{q_{12}}^{(v+u-1)},\;v=2,...,m_{12}$. Let now ${\bf r}_{lj}^{(r,v)}$ be the even image of
${\bf r}_{lj}\in s_{lj}$ in EPP$_{q_{12}}^{(v)}$ rotated by the angle $\phi_{r,v}=2\pi((r-1)p_{12}'/q_{12}+(v-1)/C)$ then its twin odd image
${\bar{\bf r}}_{lj}^{(r,v)}$ lies on ${\bar s}_{lj}\in {\bar T}_{u+v-1}$. However both the points are related by the period ${\bf D}_{lj}^{(v,u+v-1)}$
rotated by the angle $\beta_{r}=2\pi(r-1)p_{12}'/q_{12}$, i.e. we have
\be
{\bar{\bf r}}_{lj}^{(r,v)}={\bf r}_{lj}^{(r,v)}+R(\beta_r){\bf D}_{lj}^{(v,u+v-1)}
\label{A10}
\ee
where $R(\beta_r)$ is the rotation mentioned.

Therefore using the formulae \mref{Aa}, \mref{A5a} and \mref{A9} we get
\be
R(\beta_r){\bf D}_{lj}^{(v,u+v-1)}=\sum_{k=1}^{\mu_1}a_{lj,k}^{(r,v)}\alpha_k{\bf D}_1+\sum_{k=1}^{\mu_2}b_{lj,k}^{(r,v)}\beta_k{\bf D}_2
\label{A11}
\ee
and
\be
{\bf p}\cdot R(\beta_r){\bf D}_{lj}^{(v,u+v-1)}=
2\pi\ll(mZ_1C_1\sum_{k=1}^{\mu_1}a_{lj,k}^{(r,v)}\alpha_k+nZ_2C_2\sum_{k=1}^{\mu_2}b_{lj,k}^{(r,v)}\beta_k\r)\nn\\
m,n\geq \pm 1,\pm 2, ...
\label{A12}
\ee
so that
\be
\ll|\Psi_{mn}^{(12)}(x_{lj},y_{lj})\r|\leq\sum_{v=1}^{m_{12}}\sum_{r=1}^{q_{12}}\ll|e^{i{\bf p}_{mn}{\bf r}_{lj}^{(rv)}}-
                                   e^{i{\bf p}_{mn}({\bf r}_{lj}^{(rv)}+R(\beta_r){\bf D}_{lj}^{(v,u+v-1)})}\r|=\nn\\
                                   2\sum_{v=1}^{m_{12}}\sum_{r=1}^{q_{12}}\ll|\sin\ll(\fr{\bf p}_{mn}R(\beta_r){\bf D}_{lj}^{(v,u+v-1)}\r)\r|=\nn\\
2\sum_{v=1}^{m_{12}}\sum_{r=1}^{q_{12}}\ll|\sin\ll(\pi\ll(mZ_1C_1\sum_{k=1}^{\mu_1}a_{lj,k}^{(r,v)}\alpha_k+nZ_2C_2\sum_{k=1}^{\mu_2}b_{lj,k}^{(r,v)}\beta_k\r)\r)\r|=\nn\\
2\sum_{v=1}^{m_{12}}\sum_{r=1}^{q_{12}}\ll|\sin\ll(\pi\ll(m\sum_{k=1}^{\mu_1}C_1a_{lj,k}^{(r,v)}(Z_1\alpha_k-Z_{1,k})+
n\sum_{k=1}^{\mu_2}C_2b_{lj,k}^{(r,v)}(Z_2\beta_k-Z_{2,k})\r)\r)\r|\leq\nn\\
2\pi\ll(\frac{|m|I_{lj,1}}{N_1^{\frac{1}{\mu_1}}}+\frac{|n|I_{lj,2}}{N_2^{\frac{1}{\mu_2}}}\r)\nn\\
I_{lj,1}=\sum_{v=1}^{m_{12}}\sum_{r=1}^{q_{12}}\sum_{k=1}^{\mu_1}C_1|a_{lj,k}^{(r,v)}|,\;\;\;\;I_{lj,2}=\sum_{v=1}^{m_{12}}\sum_{r=1}^{q_{12}}\sum_{k=1}^{\mu_2}C_2|b_{lj,k}^{(r,v)}|\nn\\
m,n\geq \pm 1,\pm 2, ... \nn\\
\mu_1,\mu_2\leq C(\sum_{l=1}^k\sum_{j=1}^{n_l}n_{lj}-2)-m_{12}+1
\label{A13}
\ee
where $I_{lj,k},\;k=1,2,$ are finite integers while the integers $N_1,N_2$ are arbitrary. Note however that $Z_1,Z_2$ depend on $N_1,N_2$ and the larger are the latter the larger are the former and
therefore $\Psi_{mn}^{(12)}(x,y)$ describes still higher energy levels.

\subsubsection{Changing EPP}

\hskip+1,5 em  A form of EPP constructed in Sec.2.2 can be changed equivalently, i.e. by keeping a set of periods unchanged, moving any boundary RM-CPB of
EPP (i.e. an even or odd image of RM-CPB) to its another position at the EPP boundary gluing the respective boundary sides of RM-CPB making a twin parallel pair of them. A question arises how
such a change influences $\Psi_{mn}^{(12)}(x,y)$ defined on the new EPP. One expects that the respective change has to be of the same order as the order
of vanishing of $\Psi_{mn}^{(12)}(x,y)$ on the boundary of EPP given by \mref{A13}. It can be easily shown that this is the case. Namely assuming that
the twin parallel sides which the shifted RM-CPB and its other boundary image in original EPP are glued along are linked by a period ${\bf D}$ we notice
that an image point ${\bf r}_k=(x_k,y_k)$ of $(x,y)$ belonging to the RM-CPB is shifted by the same period ${\bf D}$ to its new position. Therefore we have for the new SWF
$\Psi_{mn}'(x,y)$
\be
\Psi_{mn}'(x,y)=\Psi_{mn}^{(12)}(x,y)\pm\ll(e^{i{\bf p}_{mn}{\bf r}_k}-e^{i{\bf p}_{mn}({\bf r}_k+{\bf D})}\r)
\label{A14}
\ee
from which it follows
\be
\ll|\Psi_{mn}'(x,y)-\Psi_{mn}^{(12)}(x,y)\r|\leq 2\pi\ll(\frac{|m|I_1}{N_1^{\frac{1}{\mu_1}}}+\frac{|n|I_2}{N_2^{\frac{1}{\mu_2}}}\r)\nn\\
I_1=\sum_{k=1}^{\mu_1}C_1|a_k|,\;\;\;\;I_2=\sum_{k=1}^{\mu_2}C_2|b_k|\nn\\
\label{A15}
\ee
if ${\bf D}=\sum_{k=1}^{\mu_1}a_k\alpha_k{\bf D}_1+\sum_{k=1}^{\mu_2}b_k\beta_k{\bf D}_2$ according to \mref{Aa}-\mref{Ab}.

\subsubsection{POCs in some EPP and their influence on the properties of SWF}

\hskip+1,5 em Assume that among the periods \mref{Ah} corresponding to EPP constructed in Sec.2.2 there are two ones, say ${\bf D}_1$ and ${\bf D}_2$, which
are perpendicular to each other. This condition removes any limitations on the geometry of the billiards enforced by the conditions \mref{A7} when the
skeletons chosen to quantize the billiards are parallel to the one of the periods mentioned, i.e. when the skeletons are periodic. In such a case the skeletons
can contain POCs defined by periods parallel to the one of ${\bf D}_1,{\bf D}_2$ or just by such a period itself. Each such a POC is then bounded by two
singular diagonals, i.e. two straight lines each of which must pass by at least two billiards vertex of EPP.

Let us choose $x,y$-axes to be parallel to the periods ${\bf D}_1,{\bf D}_2$ respectively. Then the singular diagonal passing by the vertex
$(x_{lj},y_{lj})$ of EPP is defined by $y=y_{lj}$ if it is parallel to the $x$-axis or by $x=x_{lj}$ if it is parallel to the $y$-one.

Consider further the case when EPP is reduced to EPP$_{q_{12}}$, i.e. $C=q_{12}$, and $q_{12}=2r_{12}$, i.e. is even. Then the distribution of momenta
passing by any point $(x,y)$ of the billiards is symmetric with respect to both the axes (see Fig.3(d)) and from \mref{A9a} we have
\be
\Psi_{mn}^{(12)}(x,y)=\nn\\
\sum_{r=1}^{r_{12}}\ll(\ll(e^{i(p_{mn,x}^{(r)}x+p_{mn,y}^{(r)}y)}+e^{-i(p_{mn,x}^{(r)}x+p_{mn,y}^{(r)}y)}\r)\r.-\nn\\
                                   \ll.\ll(e^{i(p_{mn,x}^{(r)}x-p_{mn,y}^{(r)}y)}+e^{-i(p_{mn,x}^{(r)}x-p_{mn,y}^{(r)}y)}\r)\r)=
-4\sum_{r=1}^{r_{12}}\sin\ll(p_{mn,x}^{(r)}x\r)\sin\ll(p_{mn,y}^{(r)}y\r)
\label{A16}
\ee
where according to \mref{A7a}
\be
p_{mn,x}^{(r)}=\pm\sqrt{2E_{0,m}}\cos\alpha_r+p_{n}\sin\alpha_r\nn\\
p_{mn,y}^{(r)}=\mp\sqrt{2E_{0,m}}\sin\alpha_r+p_n\cos\alpha_r\nn\\
\alpha_r=\pi(r-1)/r_{12},\;\;\;r=1,...,r_{12}
\label{A17}
\ee

Let us now rationalize periods of the considered EPP according to the following representations (see \mref{Aa}-\mref{Ab})
\be
x_{lj}\sin\alpha_r=D_1\sum_{k=1}^{\mu_1}a_{lj,k}^{(r)}X_k\nn\\
x_{lj}\cos\alpha_r=D_1\sum_{k=1}^{\mu_1}b_{lj,k}^{(r)}X_k\nn\\
y_{lj}\sin\alpha_r=D_2\sum_{k=1}^{\mu_2}c_{lj,k}^{(r)}Y_k\nn\\
y_{lj}\cos\alpha_r=D_2\sum_{k=1}^{\mu_2}d_{lj,k}^{(r)}Y_k
\label{A18}
\ee
where $a_{lj,k}^{(r)},...,d_{lj,k}^{(r)}$ are rationals while $X_k,Y_k$ are real and $D_1,D_2$ are the respective lengths of the periods ${\bf D}_1,{\bf D}_2$.

The representations \mref{A18} define according to DAT respective $C_x,C_y$ as the least common multiples for the rationals and integers $Z_x,Z_y$ for irrationals
$X_k,Y_k$ for an arbitrary given $N$ so that we have
\be
\ll|Z_xC_x\frac{x_{lj}\sin\alpha_r}{D_1}-\sum_{k=1}^{\mu_1}C_xa_{lj,k}^{(r)}Z_{x,k}\r|\leq \frac{\sum_{k=1}^{\mu_1}C_x|a_{lj,k}^{(r)}|}{N^{\frac{1}{\mu_1}}}\nn\\
\nn\\
...\;\;\;\;\;\;\;\;\;\;\;\;\;\;\;\;\;\;\;\;\;\;\;\;\nn\\
\nn\\
\ll|Z_yC_y\frac{y_{lj}\cos\alpha_r}{D_2}-\sum_{k=1}^{\mu_1}C_yd_{lj,k}^{(r)}Z_{y,k}\r|\leq \frac{\sum_{k=1}^{\mu_2}C_y|d_{lj,k}^{(r)}|}{N^{\frac{1}{\mu_2}}}
\label{A19}
\ee
where $Z_{x,k},Z_{y,k}$ are integers.

If the respective momentum ${\bf p}$ is parallel to ${\bf D}_2$, i.e. if the skeleton is periodic being parallel to ${\bf D}_2$, then the momentum and the
semiclassical correction $E_0$ to the energy $E_{mn}$ are quantized by
\be
\pm\sqrt{2E_{0,m}}D_1=2\pi mC_xZ_x\nn\\
p_nD_2=2\pi nC_yZ_y
\label{A20}
\ee

Putting now $x=x_{lj}$ in \mref{A16} we get
\be
\ll|\Psi_{mn}^{(12)}(x_{lj},y)\r|\leq \frac{\sum_{r=1}^{r_{12}}\sum_{k=1}^{\mu_1}\ll(|m|C_x|b_{lj,k}^{(r)}|+|n|C_x|a_{lj,k}^{(r)}|\r)}{N^{\frac{1}{\mu_1}}}
\label{A21}
\ee
i.e. $\Psi_{mn}^{(12)}(x,y)$ vanishes approximately on each vertical singular diagonal.

Of course similar properties $\Psi_{mn}^{(12)}(x,y)$ has on the horizontal singular diagonals.

As an illustration to the results of the above resum{\'e} we shall consider in the next two sections three examples of billiards to which these results
can be applied. These are the rectangular billiards with rectangular holes, the rectangular billiards with rotated rectangular holes and a Sinai-like
billiards \cite{3} which is the right angle triangle with the circular hole.

\section{The rectangular billiards with the rectangular holes}

\hskip+1,5 em The two rectangular billiards with the rectangular holes we are going to consider in this section are shown in Fig.6 and Fig.7.

\subsection{The rectangular billiards with the parallel rectangular holes}

\hskip+1,5 em Consider first the one of Fig.6.

Because of its geometry four angles of the outer rectangle are equal to $\pi/2$, while $4(k-1)$ of the remaining ones  - to $3\pi/2$ and sides of the
inner rectangles make angles $0$ or $\pi/2$
with the $x$-axis so that the number $C$ in Fig.2 is equal to $2$ and the genus $g$ of the multitorus is equal to $4k-3$ while
the corresponding EPP (Fig.2B) contains four images of the billiards. Each parallel pair of sides of EPP defines a period the full set of which is
exposed in Fig.2 together with their linear relations with the two of them ${\bf D}_x,{\bf D}_y$ chosen as the linear independent pair.

\begin{figure}
\begin{center}
\psfig{figure=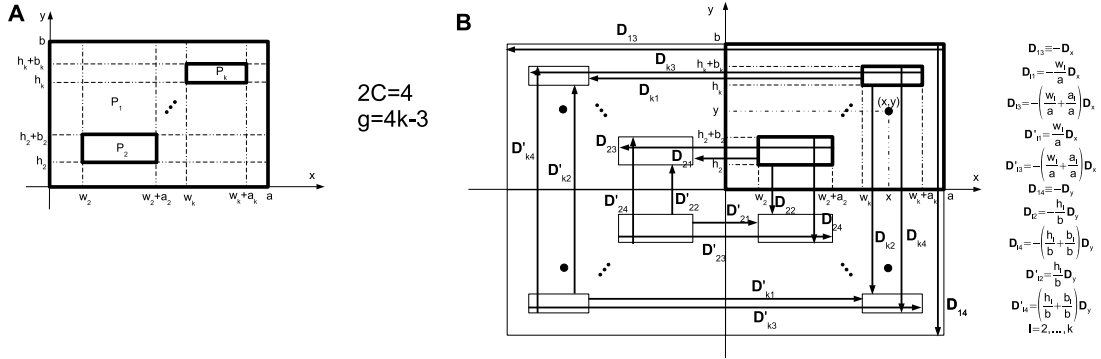,width=15 cm}
\caption{ {\bf A} - the rectangular billiards with $k-1$ parallel rectangular holes inside it, {\bf B} - its EPP. The horizontal and vertical POCs in the
billiards with their singular diagonals (SD) (dotted lines) are shown also in Fig.{\bf A}}
\end{center}
\end{figure}

According to the general rules of Sec.2 to get the semiclassical approximation of the energy spectra for the case we have first
to approximate the real coefficients present in the linear relations on Fig.2 by rational ones applying to this goal the Dirichlet theorem \mref{1}. It
is seen from the figure that there are $2(k-1)$ real numbers $w_l/a$ and $a_l/a,\;l=2,...,k,$ related to the $x$-direction which need such an
approximation as well as another $k-1$ pair of them $h_l/b$ and $b_l/b,\;l=2,...,k,$ related to the $y$-direction.
Let the corresponding integers of the Dirichlet theorem be $Z_x,N$ for the first pair and $Z_y,N$ - for the second one so that we have
\be
\ll|\frac{w_l}{a}-\frac{Z_{xw_l}}{Z_x}\r|<\frac{1}{Z_xN^\frac{1}{2(k-1)}}\nn\\
\ll|\frac{a_l}{a}-\frac{Z_{xa_l}}{Z_x}\r|<\frac{1}{Z_xN^\frac{1}{2(k-1)}}\nn\\
\ll|\frac{h_l}{b}-\frac{Z_{yh_l}}{Z_y}\r|<\frac{1}{Z_yN^\frac{1}{2(k-1)}}\nn\\
\ll|\frac{b_l}{b}-\frac{Z_{yb_l}}{Z_y}\r|<\frac{1}{Z_yN^\frac{1}{2(k-1)}}
\label{2}
\ee
where $Z_{xj},\;j=w_l,a_l$, and $Z_{yj},\;j=h_l,b_l,\;l=2,...,k$, are natural.

The quantization of the billiards on an aperiodic skeleton gives
\be
{\bf p}\cdot{\bf D}_x=2ap_x=2\pi mZ_x\nn\\
{\bf p}\cdot{\bf D}_y=2bp_y=2\pi nZ_y\nn\\
m,n=\pm 1,\pm 2,...
\label{3}
\ee
providing us with the following wave lengths on the $x,y$-directions
\be
\lambda_{xm}=\frac{2a}{|m|Z_x},\;\;\;\;\;\;\;
\lambda_{yn}=\frac{2b}{|n|Z_y}\nn\\
m,n=\pm 1,\pm 2,...
\label{3a}
\ee
which can serve as the length measure units on these directions.

Taking for example the period ${\bf D}'_{21}$ of Fig.2B we have
\be
\frac{D'_{21}}{\lambda_{xm}}=\frac{w_1}{a\lambda_{xm}}D_x=|m|Z_x\frac{w_1}{a}=|m|Z_{xw_1}+|m|\ll(Z_x\frac{w_1}{a}-Z_{xw_1}\r)
\label{3b}
\ee
so that
\be
|{D'_{21}}-|m|Z_{xw_1}\lambda_{xm}|<\frac{|m|}{N^\frac{1}{2(k-1)}}\lambda_{xm}
\label{3c}
\ee
i.e. the length of the period ${\bf D}'_{21}$ if measured in $\lambda_{xm}$-units is given by the integer $|m|Z_{xw_1}$ corrected by a small
fraction of $\lambda_{xm}$.

A set of energy levels covered approximately by this quantization is therefore given by
\be
E_{mn}=\frac{1}{2}{\bf p}^2=\frac{1}{2}\pi^2\ll(\frac{m^2Z_x^2}{a^2}+\frac{n^2Z_y^2}{b^2}\r)\nn\\
m,n=\pm 1,\pm 2,...
\label{4}
\ee

The respective semiclassical wave functions satisfying approximately the Dirichlet conditions on the billiards boundary are then given, according to
\mref{A9},
by interferences of the plane wave in all images of the point $(x,y)$ of the billiards in EPP shown in Fig.2B, i.e. we have (up to a normalization)
\be
\Psi_{mn}(x,y;Z_x,Z_y)=-\frac{1}{4}\sum_{over\;EPP}\pm e^{i{\bf p}\cdot{\bf r}_k}=\sin\ll(\pi mZ_x\frac{x}{a}\r)\sin\ll(\pi nZ_y\frac{y}{b}\r)
\label{5}
\ee
where the sum is taken over the point $(x,y)$ and its three images in EPP.

Obviously $\Psi_{mn}(x,y;Z_x,Z_y)$ vanishes on the sides $x=0, y=0, x=a$ and $y=b$ (see Fig.2A) exactly and only approximately on the remaining sides of
the billiards. In the latter cases $\Psi_{mn}(x,y;Z_x,Z_y)$ according to \mref{A13} differs from zero less than $\pi|m|N^{-\frac{1}{2(k-1)}}$ on the vertical sides and less than
$\pi|n|N^{-\frac{1}{2(k-1)}}$ on the horizontal ones. This can be shown also directly from \mref{5} by the following calculations for the side $x=w_l$
\be
|\Psi_{mn}(w_l,y;Z_x,Z_y)|=\ll|\sin\ll(\pi mZ_x\frac{w_l}{a}\r)\sin\ll(\pi nZ_y\frac{y}{b}\r)\r|\leq\ll|\sin\ll(\pi mZ_x\frac{w_l}{a}\r)\r|=\nn\\
                          \ll|\sin\ll(\pi mZ_x\ll(\frac{w_l}{a}-\frac{Z_{xw_l}}{Z_x}\r)\r)\r|<\frac{|m|\pi}{N^\frac{1}{2(k-1)}}
\label{6}
\ee

Therefore $\Psi_{mn}(x,y;Z_x,Z_y)$ and its corresponding energy levels will approximate well the exact states and energies rather for
the high energy region, i.e. for large $Z_x,Z_y$ and $N$.

One can notice that the form \mref{5} of SWF and its energy spectrum \mref{4} both coincide with the ones of the respective rectangle with the sides
$a/Z_x,b/Z_y$ and without a hole inside, i.e. in the considered case the exact wave functions of the billiards and its high energy spectra are well
approximated by the respective quantities of the rectangular billiards mentioned. Obviously the respective approximation with a desired accuracy
determined by the number $N$ is possible due to the special tuning of the numbers $Z_x,Z_y$ provided by DAT.

To understand the numerical properties of SWF \mref{5} it is necessary to consider skeletons which are parallel to some periods. They are important
because of POCs which must then appear in such skeletons together with their singular diagonals. According to Sec.2.5.5 we can consider periodic skeletons
which do not put any constraints on the form of the billiards itself and which can be identified by momenta directed along the $x$- or $y$-axis.

According to Sec.2.5.5 the numbers $Z_x,Z_y$ in \mref{2} are the same as the ones determining the inequalities \mref{A19}. Therefore on the singular
diagonals corresponding to POCs defined by the periods of EPP of Fig.6B parallel to the $x$- or $y$-axis and shown schematically on Fig.6A SWF
$\Psi_{mn}(x,y;Z_x,Z_y)$ must vanish approximately according to \mref{A21}. In the considered case of the billiards this coincides of course with
its approximate vanishing on the inner boundaries of the billiards.

In other words the existence of POCs in periodic skeletons is already hidden in the formula \mref{5} proving its immanent composition of POCs. The
latter fact imprints itself by singular diagonals being the boundaries of POCs along which SWF \mref{5} vanishes approximately according to \mref{A21}.
These approximate nodal lines of $\Psi_{mn}(x,y;Z_x,Z_y)$ are the main property of the latter demonstrating its structure as being composed of POCs \cite{53}.

Let us notice further that the choice of the particular periods ${\bf D}_x,{\bf D}_y$ as the base on the plane was quite arbitrary and in fact can be
substituted by any other pair of the linear independent periods taken from the set of them shown in Fig.2B. Such another choice causes changes of the
real coefficients and the respective numbers $Z_x,Z_y$ in \mref{2} so that the formulae \mref{4} as well as the corresponding SWFs \mref{5} cover then
different energy regions of the problem according to the point 2 of Sec.2.5.1. Nevertheless such a change in no way modifies the running of SDs in Fig.2A and Fig.3.

\subsection{The rectangular billiards with rotated rectangular holes}

\hskip+1,5 em The billiards is shown in Fig.7A. A rotation of the $k-1$ inner rectangles with respect to the $x$-axes by $\pi/4$ changes the number $C$, i.e.
the least common multiple of the angle denominators, by two so that $C=4$ now in comparison with the previous case. The corresponding EPP build
according to the recipe of Sec.2.2 is then given on Fig.4B. Identifying its parallel boundary sides EPP is then changed into the closed two dimensional
surface with genus $g=8k-7$ and therefore having $16k-14$ independent (in the space of integers) periods which can be found among $16k-10$ of them
linking twin parallel sides. However on the real plane of the respective RM-CPBRS there are only two periods which can be
independent chosen as ${\bf D}_x\equiv {\bf D}_{14}^{22}$ and ${\bf D}_y\equiv -{\bf D}_{14}^{11}$. The linear relations of the remaining ones with
the two chosen can be determined according to Sec.2.3, i.e. by differences of the respective coordinates of vertices of EPP and their rationalization
can be performed as in Sec.2.5.5 and the results are shown in Fig7D. Therefore applying DAT we have according to Fig.7D

\begin{figure}
\begin{center}
\psfig{figure=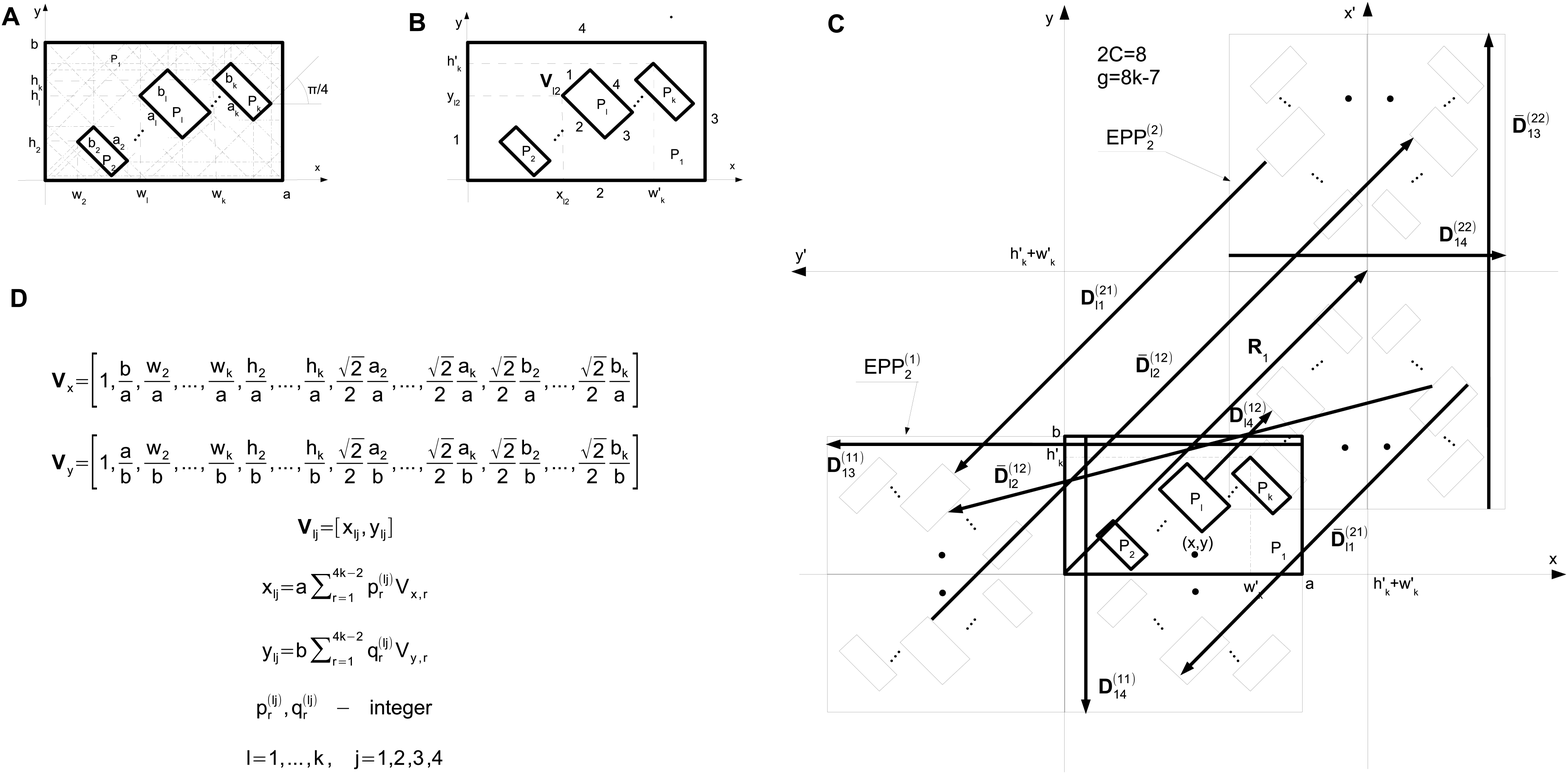,width=15 cm}
\caption{{\bf A} - the rectangular billiards with $k-1$ rectangular holes inside it rotated by the angle $\pi/4$; {\bf B} - enumeration of sides of the
billiards; {\bf C} - its EPP; {\bf D} - the coordinates of vertexes of the billiards. Fig.A shows also (schematically) the system of POCs with their
singular diagonals. These SDs are the approximate nodal lines of
$\Psi_{mn}(x,y;Z_x,Z_y)$}
\end{center}
\end{figure}

\be
\ll|V_{x,r}-\frac{Z_{xr}}{Z_x}\r|<\frac{1}{Z_xN^\frac{1}{4k-3}}\nn\\
\ll|V_{y,r}-\frac{Z_{yr}}{Z_y}\r|<\frac{1}{Z_yN^\frac{1}{4k-3}}
\label{7}
\ee

The next steps are rather standard ones.

First quantizing on an aperiodic skeleton along which a plane SWF is propagating with a momentum {\bf p} we get
\be
{\bf p}\cdot{\bf D}_x=2ap_x=2\pi mZ_x\nn\\
{\bf p}\cdot{\bf D}_y=2bp_y=2\pi nZ_y\nn\\
m,n=\pm 1,\pm 2,...
\label{8}
\ee
and for the respective energy levels we have
\be
E_{mn}=\frac{1}{2}{\bf p}^2=\frac{1}{2}\pi^2\ll(\frac{m^2Z_x^2}{a^2}+\frac{n^2Z_y^2}{b^2}\r)\nn\\
m,n=\pm 1,\pm 2,...
\label{9}
\ee

SWFs corresponding to \mref{9} and constructing according to Sec.2.5.2 is given by
\be
\Psi_{mn}(x,y;Z_x,Z_y)=\nn\\
                        \sin\ll(\pi mZ_x\frac{x}{a}\r)\sin\ll(\pi nZ_y\frac{y}{b}\r)-
                       e^{i\pi\ll(m\frac{Z_x}{a}+n\frac{Z_y}{b}\r)(h'+w')}\sin\ll(\pi mZ_x\frac{y}{a}\r)\sin\ll(\pi nZ_y\frac{x}{b}\r)
\label{10}
\ee

Obviously each $\Psi_{mn}(x,y;Z_x,Z_y)$ vanishes exactly on the sides $x=0$ and $y=0$ of the outer rectangle but
only approximately on the remaining sides of the billiards.

The POC structure of EPP is again important for understanding the properties of SWFs \mref{10}. There are at least two pairs of periods orthogonal to
each other in each pair - the one already considered, i.e. ${\bf D}_x,{\bf D}_y$ and the other one rotated by the angle $\pi/4$ with respect to the
former. Both the pairs of periods generate respective POCs which SDs are shown schematically on Fig.7A and some of which coincide partly with respective
sides of the billiards. According to Sec.2.5.5 SWF \mref{10} vanishes approximately along these SDs.

\begin{figure}
\begin{center}
\psfig{figure=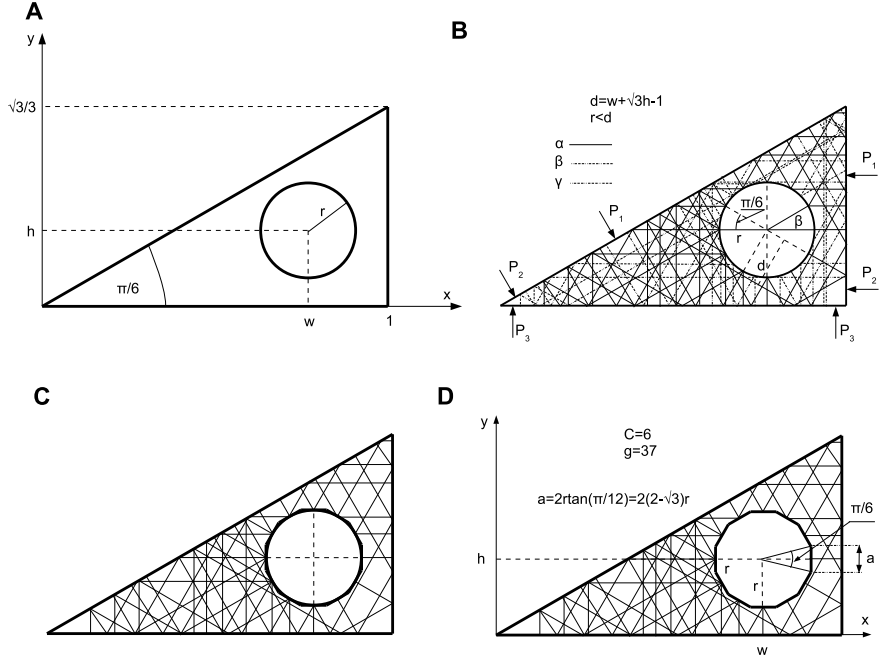,width=12 cm}
\caption{{\bf A} - the Sinai-like billiards; {\bf B} - its $31$ shortest periodic orbits the reflecting points of which on the circular boundary are
uniformly distributed around it except the $\beta$-sector; $P_1$, $P_2$ and $P_3$ denote the three superscar POCs in the billiards; {\bf C} - its
approximation by RM-CPB determined by $21$ periodic orbits uniformly distributed around the circular boundary;
{\bf D} - the resulting RM-CPB with the regular dodecagon substituting the circular boundary}
\end{center}
\end{figure}

\section{The Sinai-like billiards \cite{3}}

\subsection{Semiclassical quantization of the Sinai-like billiards}

\hskip+1,5 em The billiards is shown in Fig.8A. It is the right triangle with one of its acute angles equal to $\pi/6$ and with a circular hole in it. To
quantize it semiclassically we have used the idea formulated in our earlier paper \cite{54} that
each billiards can be identified by the set of its all periodic orbits and can be approximated by a subset of its shortest ones. This idea allows us to
approximate any billiards with holes inside it by a polygon one with polygon holes. Each side of the approximating billiards is taken to be tangent to the
approximated one at a point where some of the shortest periodic orbit of the approximated billiards is reflecting by the billiards boundary. Considering
the Sinai-like billiards shown in Fig.8A its respective approximation done according to this description is shown in Fig.8B,C,D where $31$ of the shortest
periodic orbits used to the construction of this approximation are also drawn.
These $31$ periodic orbits would allow us to envelope the circular inner boundary of the billiards by $23$-side polygon billiards but the lack
of short periodic orbits emerging (or reflecting) from the points of the circle defined by the angle $\beta$ on Fig.8B prevents the followed approximations
to be better than the ones got by enveloping the circle by the regular dodecagon shown in Fig.8C, i.e. by using only $21$ of the shortest periodic
orbits shown in Fig.8C. Therefore limiting to these orbits we get additionally two bonuses of such an approximation of the Sinai-like billiards given by
Fig.8A - 1. the rational billiards and 2. the corresponding EPP can be constructed as a single leaf on a plane. Nevertheless in opposite case,
i.e. when some of angles of an approximating billiards appeared irrational, it would have necessary first to be rationalized by approximating its
angles by rational ones applying DAT (see Introduction).

EPP for the approximating billiards Fig.8D is shown in Fig.9A with its some periods the full number of which is equal to $78$ among which there are $74$
linearly independent in the space of integers. The linear relations shown in the table {\bf B} in Fig.9 are written for the periods defined by the
original billiards numbered as $1$ in Fig9A while the remaining ones determined by the images $2,...,6$ of the billiards can be obtained by the table
{\bf B} by substituting there the pair of periods $({\bf D}_1,{\bf D}_2)$ by the pairs $({\bf D}_1(\pi/3),{\bf D}_2(\pi/3))$,
$({\bf D}_1(2\pi/3),{\bf D}_2(2\pi/3)$, ... , $({\bf D}_1(5\pi/3),{\bf D}_2(5\pi/3))$ respectively and using their relations with ${\bf D}_1$ and
${\bf D}_2$ shown in the table {\bf C} of Fig.9.

Inspecting
further these linear relations one can notice that the coefficients by which all these periods are related with the period ${\bf D}_1$ and ${\bf D}_2$ are
linear dependent on the following four real numbers $\{w,\sqrt{3}h,r,\sqrt{3}r\}$ with coefficients which are fractions with the least
common denominator equal to six.

\begin{figure}
\begin{center}
\psfig{figure=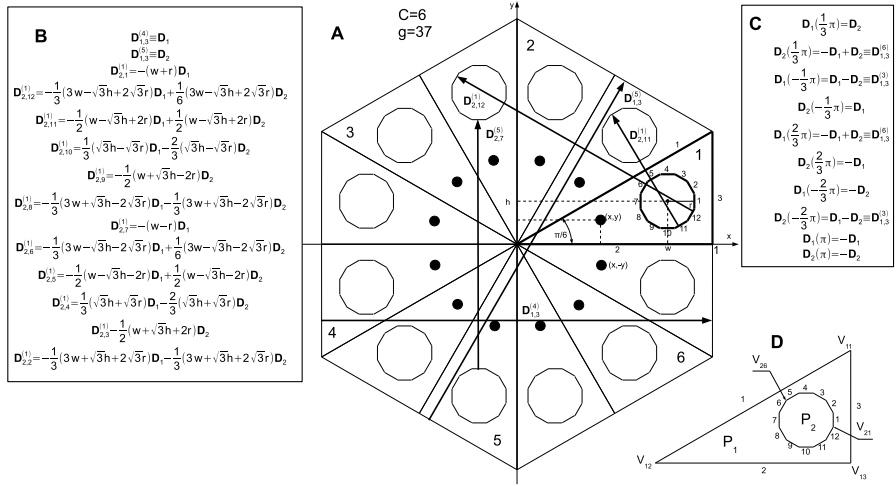,width=12 cm}
\caption{{\bf A} - EPP corresponding to RM-CPB approximating the Sinai-like billiards and some of its $78$ periods. The periods shown are marked
according to Sec.2.2; {\bf B} - the table of the linear relations between the periods defined by the sides of the original RM-CPB; {\bf C} - the table
defining five other pairs of independent periods substituting the ones ${\bf D}_1,{\bf D}_2$ in the table {\bf A} to get the linear relations for the
remaining periods determined by the images $2,...,6$ of the original RM-CPB; {\bf D} - the enumeration of sides and vertices in RM-CPB approximating the
Sinai-like billiards}
\end{center}
\end{figure}

Applying therefore the Dirichlet theorem we can approximate this set of real numbers by respective rationals as follows
\be
\ll|x_k-\frac{Z_{k}}{Z}\r|<\frac{1}{ZN^\frac{1}{4}}
\label{12}
\ee
where $x_k,\;k=1,...,4$, denote the real numbers of the set.

The next steps are standard.

Quantizing the classical momentum {\bf p} on an aperiodic skeleton we get
\be
{\bf p}\cdot{\bf D}_1=12\pi mZ\nn\\
{\bf p}\cdot{\bf D}_2=12\pi nZ\nn\\
m,n=\pm 1,\pm 2,...
\label{13}
\ee
and according to \mref{A6a}
\be
E_{mn}=\frac{1}{2}{\bf p}^2=96\pi^2Z^2(m^2-mn+n^2)\nn\\
m,n=\pm 1,\pm 2,...
\label{14}
\ee
for the respective energy levels.

SWFs corresponding to \mref{13} and \mref{14} are
\be
\Psi_{mn}(x,y;Z)=\sin(6\pi mZx)\sin(2\pi\sqrt{3}(2n-m)By)+\nn\\
                       \sin(3\pi mZ(x-\sqrt{3}y))\sin(\pi Z\sqrt{3}(2n-m)(\sqrt{3}x+y))-\nn\\
                        \sin(3\pi mZ(x+\sqrt{3}y))\sin(\pi Z\sqrt{3}(2n-m)(\sqrt{3}x-y))\nn\\
                        m=1,2,...,\;n=\pm 1,\pm 2,...
\label{15}
\ee
as a result of applying the formulae \mref{A9}-\mref{A10} of Sec.2.5.1.

According to \mref{A9a} the latter formula can be rewritten for a later discussion on POCs as follows
\be
\Psi_{mn}(x,y;Z)=\sin(6\pi mZx)\sin(2\pi\sqrt{3}(2n-m)By)-\nn\\
                       \sin(6\pi nZx)\sin(2\pi\sqrt{3}Z(n-2m)y)+\nn\\
                       \sin(6\pi (n-m)Zx)\sin(2\pi\sqrt{3}Z(n+m)y)
\label{15a}
\ee

In fact $\Psi_{mn}(x,y;Z)$ in \mref{15}-\mref{15a} provide us with the exact solutions for the triangular billiards, i.e. with the internal circular boundary
removed. Therefore it has to vanish exactly on the sides $x=1$, $y=0$ and $y=\sqrt{3}x$ of the triangle and as it follows from Sec.2.5.3 only
approximately on the remaining sides of RM-CPB of Fig.8D.

As it follows from Sec.2.5.5 the quantization on the periodic skeletons as the ones shown on Fig.10A adds the states with the quantum numbers
$m=0,\;n\geq 1$ and
$m\geq 1,\;n=0$. Fig.10A shows also runnings of singular diagonals on which $\Psi_{mn}(x,y;Z)$ approximately vanishes demonstrating in this way the existence
of POCs which EPP for the case considered is composed of.

It can be checked also that $\Psi_{mn}(x,y;Z_x,Z_y)$ vanishes approximately on each unstable periodic orbits shown in Fig.8B, i.e. are limited
by inequalities similar to \mref{A21}. For example, putting $x=w$ in \mref{15a} we get
\be
\ll|\Psi_{mn}(w,y;Z)\r|<6\pi\frac{m+|n|+|m-n|}{N^\frac{1}{4}}
\label{17}
\ee

\begin{figure}
\begin{center}
\psfig{figure=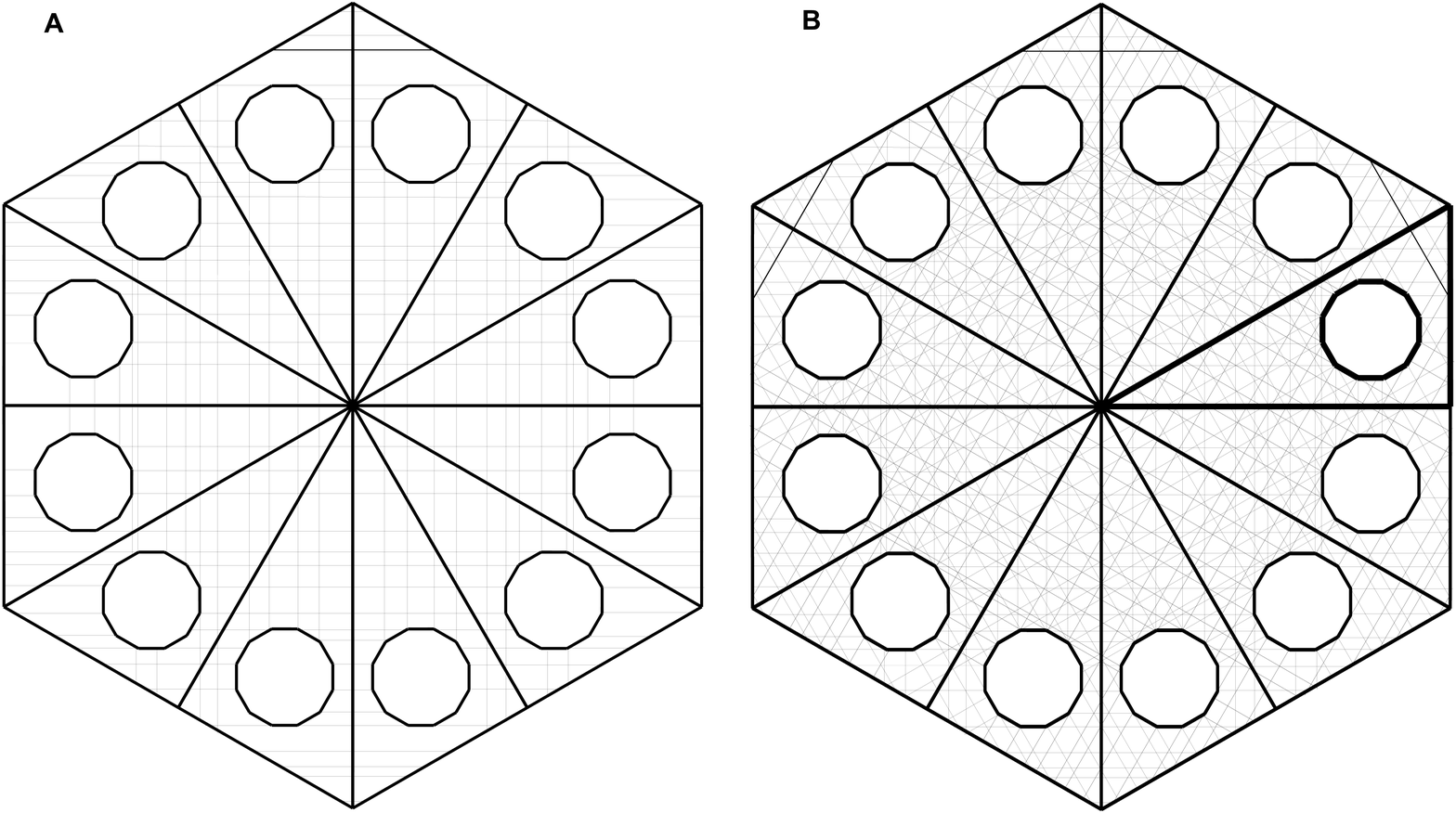,width=14 cm}
\caption{{\bf A} - a net of SDs of horizontal and vertical POCs in EPP for RM-CPB approximating the Sinai-like billiards. The net corresponds to the
pair of the orthogonal periods $({\bf D}_1,-{\bf D}_1+2{\bf D}_2)$ - only those POCs are shown which are generated by the periodic orbits of the
Sinai-like billiards. {\bf B} - the join net of SDs corresponding to the orthogonal pairs of periods $({\bf D}_1,-{\bf D}_1+2{\bf D}_2)$,
$({\bf D}_2,-{\bf D}_1+2{\bf D}_2)$ and $({\bf D}_1-{\bf D}_2,{\bf D}_1+{\bf D}_2)$. $\Psi_{mn}(x,y;Z_x,Z_y)$ vanishes approximately on this net.}
\end{center}
\end{figure}

\subsection{Scars and superscars in the Sinai-like billiards}

\hskip+1,5 em In the semiclassical approximation of the Sinai-like billiards as given in Sec.4.1 above particularly when its quantization is performed on the
periodic skeletons shown in Fig.10A the presence of POCs is as natural as typical for the periodic skeletons in RM-CPB. However in this context an interesting
question arises what happens to POCs when still new longer periodic orbits of the Sinai-like billiards are included into the considerations to make the respective semiclassical
approximation still more accurate. It is clear that adding new periodic orbits will change the form of polygons enveloping the circular boundary of the
Sinai-like billiards which will be no longer regular one and will provide us with new POCs defined by the new orbits. Nevertheless a general rule is
that POCs defined by an old set of shorter periodic orbits survive changing possibly their sizes. As a rule POCs defined by a set of stable periodic
orbits forming superscars in the Sinai-like billiards remain almost unchanged
while the sizes of the ones defined by isolate stable or unstable periodic orbits of the Sinai-like billiards are being shrunk by subsequently added
orbits
gradually decreasing with increasing number of new periodic orbits included. Note however that the zero limit of breadths of the latter POCs
cannot be achieved just because
the semiclassical approximations are asymptotic in principle which is reflected by higher and higher energies to be considered if one wants to get more
and more accurate results. Simultaneously however since lengths of the new POCs increase then runnings of the latter through the billiards when folded become
still more complicated and tangled up. As a result of such a behaviour a given point of the billiards can be multiply covered by a single long POC passing
by it from many different and "random" directions. Therefore SWFs $\Psi_{mn}(x,y;Z_x,Z_y)$ defined in the point considered appears as a result of multiple
superpositions of BSWF propagating along the POC with itself and with many different phases gained on the ways of its propagation along the POC between
two subsequent passes through the point.
Simultaneously there are other long POCs contributing to $\Psi_{mn}(x,y;Z_x,Z_y)$ at the point considered in a similar way. Therefore the longer
periodic orbits are included into the considerations the more "chaotic" becomes figure of forming $\Psi_{mn}(x,y;Z_x,Z_y)$ in the Sinai-like billiards.
Fig.10B illustrate to some extent this situation.

\section{Summary and conclusions}

\hskip+1,5 em In this paper we have extended our approach to semiclassical quantization of the polygon billiards formulated and used in our earlier
papers \cite{41}- \cite{54} onto such billiards with polygon holes, i.e. which are multiconnected. The basic principles of our approach have been
formulated in the papers mentioned. However in the present paper they are extended, ordered and completed by several conclusions of a general meaning
which in the previous paper have been mentioned only occasionally. All these have been done in Sec.2. Let us therefore enumerate the points of the latter section most
important for the present paper. They are
\begin{enumerate}
\item the standardization of the construction of EPP for a general rational multi-connected polygon billiards (Sec.2.2);
\item the standardization of finding of all important periods of EPP (Sec.2.2);
\item the standardization of rationalizing of linear relations between periods of EPP (Sec.2.3); and
\item the standardization of construction of SWF on EPP (Sec.2.5).
\end{enumerate}

It is also necessary to mention the role of DAT discussed widely in Introduction. For the quantum physics in billiards it establishes definitely that
\begin{itemize}
\item there are always the maximal wave lengths which can be used as length units in billiards and which allow us to measure lengths of the shortest
periodic orbits, essential for the semiclassical quantization of billiards, with desired accuracies;
\item the better the accuracies are to be the shorter these maximal wave lengths must be and therefore the higher are energy levels approximated
semiclassically;
\item as it was discussed in Introduction in the case of irrational multi-connected polygon billiards it allows us to substitute such billiards by the
rational ones with desired accuracy;
\end{itemize}

The extension done in the paper allowed us then to apply our semiclassical approach to billiards which are to some extent arbitrary, i.e. which
boundaries both the outer and the inner ones are curved. In the last case we have considered in Sec.4 the Sinai-like billiards applying the procedure
first used in our earlier paper \cite{54} to the Bunimovich stadium which approximates the curved boundaries helping with shortest periodic orbits of the
considered billiards. Such an approach allowed us to incorporate periodic orbits of the original non-RM-CPB billiards into a set of such orbits of RM-CPB
approximated the original one. As it was observed SWFs constructed for the Sinai-like billiards possessed the property of vanishing approximately on periodic
orbits used in approximating the billiards by the polygon one. Such a property was observed also in the case of the Bunimovich stadium considered in our
earlier paper \cite{54} which the property has been called there the anty-scar one as the opposite to the scar phenomenon of Heller \cite{44}. However it is not clear that this property is maintained when the
longer periods of the billiards are included to construct their approximations by the (multi-connected) polygons.

Let us note finally that although it was not done in the present paper an accuracy of such approximations of non-RM-CPB can be estimated (see
\cite{54}) to show their agreement with the demands of the respective mathematical theorems \cite{40}.

\end{document}